# Symmetry-Based Transport Criterion and Robust Design for $RuO_2$-type Altermagnets

Zhaoqing Li,[1] Yi Liu,[2,3,*] Yizheng Wu,[4] Zhe Yuan[1,†]

[1]*State Key Laboratory of Surface Physics and Interdisciplinary Center for Theoretical Physics and Information Sciences, Fudan University, Shanghai 200433, China*

[2]*Institute for Quantum Science and Technology, Shanghai University, Shanghai 200444, China*

[3]*Department of Physics, Shanghai University, Shanghai 200444, China*

[4]*State Key Laboratory of Surface Physics and Department of Physics, Fudan University, Shanghai 200433, China*

[*] yiliu42@shu.edu.cn; [†] yuanz@fudan.edu.cn

**Abstract:**

Altermagnetism combines vanishing net magnetization with momentum-dependent spin splitting, enabling spin-transport functionalities without ferromagnetism. Although rutile $RuO_2$ has been proposed as a *d*-wave altermagnet, its magnetic ground state remains highly controversial. Moreover, existing transport verifications suffer from severe ambiguity due to the entanglement of spin-splitting effects and time-reversal-even spin-Hall contributions. Here, we introduce a two-pronged strategy to break the current verification impasse. First, we establish a definitive, symmetry-based transport criterion: by identifying specific crystallographic geometries where the spin-Hall background is strictly forbidden, we provide a zero-background window for unambiguous altermagnetic detection. Concurrently, to bypass the uncertain bulk energetics, we propose the $RuO_2/MnO_2$ superlattice as a prototypical platform to robustly enforce the *d*-wave altermagnetic order via interfacial magnetic proximity. Within this platform, symmetry-resolved spin-transport analysis reveals unconventional responses, including collinear spin currents enabling field-free magnetization switching and sizable spin-polarized charge currents. By combining a universal transport criterion with a robustly engineered material platform, our work transcends the debates over bulk properties, providing an actionable, definitive experimental roadmap for verifying *d*-wave altermagnetism.

*Introduction.*—Altermagnetism has recently emerged as a distinct magnetic order, fundamentally different from conventional ferromagnetism and antiferromagnetism [1,2]. Within spin group theory [3-10], altermagnets are characterized by spin sublattices related by real-space rotations or mirror operations [1-2], rather than by translations or inversion. Consequently, altermagnets exhibit vanishing net magnetization, as in antiferromagnets, while simultaneously hosting momentum-dependent spin splitting in their energy bands [11-14], a hallmark traditionally associated with ferromagnets. This unusual coexistence combines the absence of stray fields and ultrafast spin dynamics of antiferromagnets with nonrelativistic spin splitting, positioning altermagnets as an attractive platform for spintronic applications [15-17].

Depending on the symmetry of the spin texture, altermagnets can be classified into even-parity types including *d*-, *g*-, and *i*-wave orders [1,2], whose Fermi surfaces host two, four, and six nodes, respectively. Among these, the *d*-wave altermagnets are particularly compelling. Their extreme momentum-space anisotropy gives rise to a nonrelativistic spin-charge conversion mechanism, enabling spin-polarized transport and giant spin currents without spin-orbit coupling (SOC) [11,18-20]. These unconventional transport properties make *d*-wave altermagnets a promising setting for both fundamental studies and high-performance spintronic devices.

Rutile $RuO_2$ has been proposed as a prototypical *d*-wave altermagnet, owing to its simple crystal structure and predicted large spin splitting. Early neutron diffraction [21] and resonant x-ray scattering experiments [22] suggested an antiferromagnetic ground state. Subsequent angle-resolved photoemission spectroscopy (ARPES) reported giant spin splitting and directly visualized *d*-wave spin texture [23], while magnetic circular dichroism provided evidence of time-reversal symmetry breaking in the band structure [24]. Transport experiments further revealed anomalous Hall responses [25-26] and spin-current generation [11,18,27-33], broadly consistent with altermagnetism.

However, this picture has been challenged by recent studies. Muon spin rotation measurements detected negligible local magnetic moments in $RuO_2$ [34,35], and later ARPES measurements observed electronic structures consistent with a nonmagnetic state [36]. These findings are supported by optical conductivity [37] and quantum oscillation measurements [38], which found no signature of altermagnetic order. Moreover, spin-charge conversion effects initially attributed to inverse spin splitting have been reinterpreted in terms of an anisotropic inverse spin Hall effect [39]. Exacerbating this controversy, most recently, spin Seebeck measurements conclusively reported the entire absence of transport altermagnetic spin-splitting effects in epitaxial $RuO_2$ films [40]. In typical low-symmetry orientations, the time-reversal-odd (T-odd) spin-splitting effect (SSE) and the time-reversal-even (T-even) anisotropic spin Hall effect (ASHE) share the same angular dependencies [18,39]. Consequently, previous attempts to disentangle these effects relying on reference-based subtraction [33] or empirical scaling [18,30-31] have proven inadequate, leaving the true ground state of $RuO_2$ and its suitability for spin-splitting-based applications under active debate.

In this Letter, we introduce a two-pronged strategy to break the current impasse surrounding $RuO_2$-type *d*-wave altermagnetism. To overcome the critical bottleneck of transport entanglement, we establish a definitive, symmetry-based transport criterion. By identifying specific crystallographic geometries where the T-even spin Hall background is strictly symmetry-forbidden, we provide a true zero-background window to unambiguously isolate the pure T-odd altermagnetic responses without relying on phenomenological subtraction. Concurrently, rather than relying on the highly controversial and fragile energetics of bulk $RuO_2$, we propose superlattice engineering. By employing a $RuO_2$/$MnO_2$ superlattice as a prototypical platform, we robustly enforce the *d*-wave altermagnetic order via interfacial magnetic proximity. Within this robust platform, we predict distinctive spin-transport properties, including collinear spin currents enabling field-free switching of perpendicular magnetization and spin-polarized currents yielding large tunneling magnetoresistance (TMR). By combining a background-free transport criterion with a stabilized material platform, our work shifts the paradigm from debating fragile bulk energetics to providing an actionable, definitive experimental roadmap for verifying *d*-wave altermagnetism.

*Robust altermagnetic platform via superlattice engineering.*—To provide a controlled platform for testing our symmetry-based transport criterion, we consider a $RuO_2$/$MnO_2$ superlattice composed of alternating rutile layers [Fig. 1(a)]. Both $RuO_2$ and $MnO_2$ crystallize in the rutile-type tetragonal structure (space group $P4_2/mnm$), in which transition-metal ions occupy the centers of slightly elongated oxygen octahedra [Fig. 1(b)]. These octahedra are related by real-space rotations rather than inversion or translation, providing the crystallographic environment required for momentum-dependent magnetic behavior. The two constituents also have identical bonding topology and closely matched in-plane lattice constants, with a lattice mismatch of only 1.8%, ensuring favorable structural compatibility for epitaxial integration. Experimentally demonstrated rutile-oxide superlattices, such as $TiO_2/VO_2$ [41], further establish a precedent for controlled layering of rutile oxides. Our first-principles calculations provide additional support for the stability of the proposed $RuO_2/MnO_2$ structures, yielding a negative formation energy and the absence of imaginary phonon modes [42]. These structural and energetic considerations, combined with the robust antiferromagnetic order of $MnO_2$, motivate $RuO_2/MnO_2$ as a suitable platform in which interfacial exchange coupling can stabilize the otherwise correlation-sensitive magnetic order of $RuO_2$.

To contrast the robust altermagnetism of our engineered superlattice with the fragile nature of bulk $RuO_2$, we perform comparative first-principles calculations. Figure 1(c) shows the calculated magnetic moment on one Ru sublattice as a function of the effective Hubbard interaction $U - J$ applied to the Ru 4$d$ orbitals, while the other sublattice hosts the precisely opposite moment forming the altermagnetic order. In bulk $RuO_2$, the moment on each Ru site remains negligible for $U - J < 1.2$ eV, and increases rapidly at larger values, reflecting the strong sensitivity to correlation parameters reported previously [52,53] and the resulting ambiguity in determining its magnetic ground state. In striking contrast, the $RuO_2/MnO_2$ superlattice exhibits sizable local Ru moments across the entire range of $U - J$ considered. These moments form a collinear antiferromagnetic order, stabilized by an altermagnetic proximity effect [54] via interfacial exchange coupling to antiferromagnetic $MnO_2$. Specifically, the $MnO_2$ template injects a staggered effective Zeeman field that strongly pins the induced Ru spins, as evidenced by their spatial decay away from the interface. This robust interfacial pinning fundamentally suppresses any spontaneous in-plane magnetic reconstruction [42]. Notably, the vanishing net magnetization is not due to compensation between the different Ru and Mn species, but rather because each constituent layer independently satisfies antiferromagnetic compensation guaranteed by the magnetic group symmetry [14]. Including SOC produces only minor quantitative changes, confirming Ru magnetism is predominantly driven by exchange. To further assess the energetic stability of this magnetic ordering, we compared its total energy with those of four representative competing antiferromagnetic states. The altermagnetic state is energetically favored over all four states for *n*=1–4, with appreciable energy differences throughout the investigated range [42].

The calculated electronic band structure of the $RuO_2/MnO_2$ superlattice is shown in Fig. 1(d). Pronounced spin splitting appears along the high-symmetry paths M-Γ and A-Z in the absence of SOC. This specific splitting pattern is strictly dictated by the magnetic space group $P\text{-}4'2_1m'$ (No. 113.270) of the superlattice. Under this symmetry, the effective altermagnetic exchange field is constrained to the $d_{xy}$-type form, $h_{ex}(\boldsymbol{k}) \propto k_x k_y \sigma_z$. Consequently, the exchange field vanishes along the principal axes and planes with $k_x = 0$ or $k_y = 0$ (e.g., the Γ-X, Γ-Z, and Z-R paths), enforcing spin degeneracy, while it reaches its extrema along the diagonal M-Γ and A-Z paths. Figure 1(e) shows the first Brillouin zone and Fig. 1(f) the spin-resolved Fermi surface contours at two representative $k_z$ planes. In both the $k_z = 0$ and $k_z = \pi/c$ planes, the spin-up (red) and spin-down (blue) Fermi surface contours are related by a 90° rotation. This characteristic spin splitting in momentum space, coexisting with the collinear antiferromagnetic order in real space, constitutes a defining fingerprint of altermagnetism in the $RuO_2/MnO_2$ superlattice. The same characteristic *d*-wave spin splitting pattern persists across the investigated superlattice periods [42].

*The entanglement of spin-transport responses in conventional geometries.*— With the *d*-wave altermagnetic order firmly established in our prototypical superlattice, we now utilize this platform to systematically examine the fundamental challenge in spin-transport measurements: the entanglement of spin-splitting effects with spin Hall backgrounds. We evaluate the spin conductivity tensor $\sigma_{\alpha,\beta}^{\gamma}$ within linear-response theory using the Kubo formalism [11,55-57]. Here $\alpha$, $\beta$, and $\gamma$ denote the directions of spin-current flow, applied electric field and spin polarization, respectively. The total spin conductivity can be decomposed into two symmetry-distinct components: a T-odd component, originating from spin splitting in the absence of SOC and modified by the magnetic SHE (MSHE) when SOC is included [11,57-59], and a T-even component arising from the conventional SHE. These contributions are given by

$$\sigma_{\alpha,\beta}^{\mathrm{odd},\gamma} = -\frac{e\hbar}{\pi N_k V}\sum_{\boldsymbol{k},n,m}\frac{\eta^2 \mathrm{Re}\left[\langle n\boldsymbol{k}|\hat{j}_\alpha^\gamma|m\boldsymbol{k}\rangle\langle m\boldsymbol{k}|\hat{v}_\beta|n\boldsymbol{k}\rangle\right]}{[(\varepsilon_F-\varepsilon_{\boldsymbol{k}n})^2+\eta^2][(\varepsilon_F-\varepsilon_{\boldsymbol{k}m})^2+\eta^2]}, \quad (1)$$

$$\sigma_{\alpha,\beta}^{\mathrm{even},\gamma} = -\frac{e\hbar}{2\pi N_k V}\sum_{\boldsymbol{k},n\neq m}\mathrm{Im}\left[\langle n\boldsymbol{k}|\hat{j}_\alpha^\gamma|m\boldsymbol{k}\rangle\langle m\boldsymbol{k}|\hat{v}_\beta|n\boldsymbol{k}\rangle\right]$$
$$\times\left\{\frac{\eta\,(\varepsilon_{\boldsymbol{k}m}-\varepsilon_{\boldsymbol{k}n})}{[(\varepsilon_F-\varepsilon_{\boldsymbol{k}n})^2+\eta^2][(\varepsilon_F-\varepsilon_{\boldsymbol{k}m})^2+\eta^2]}\right.$$
$$+\frac{2\eta}{(\varepsilon_{\boldsymbol{k}n}-\varepsilon_{\boldsymbol{k}m})[(\varepsilon_F-\varepsilon_{\boldsymbol{k}m})^2+\eta^2]}$$
$$\left.+\frac{2}{(\varepsilon_{\boldsymbol{k}n}-\varepsilon_{\boldsymbol{k}m})^2}\mathrm{Im}\left(\ln\frac{\varepsilon_{\boldsymbol{k}m}-\varepsilon_F-i\eta}{\varepsilon_{\boldsymbol{k}n}-\varepsilon_F-i\eta}\right)\right\}, \quad (2)$$

where $|n\boldsymbol{k}\rangle$ and $\varepsilon_{\boldsymbol{k}n}$ are the Bloch states and energies, $\varepsilon_F$ is the Fermi energy, $\hat{v}_\beta$ is the velocity operator, and $\hat{j}_\alpha^\gamma$ is the spin-current operator with propagating direction $\alpha$ and polarization along $\gamma$. The broadening parameter $\eta$ models disorder-induced scattering and is related to the relaxation time $\tau$ via $\tau = \hbar/2\eta$. In the clean limit $\eta \to 0$, the T-odd contribution scales as $1/\eta$, analogous to the electrical conductivity in the relaxation-time approximation, whereas the T-even contribution remains unchanged and corresponds to the intrinsic spin Hall response.

We begin our analysis with the conventional principal-axis configuration, i.e. $x$‖[100], $y$‖[010], and $z$‖[001]. In this geometry, the crystal symmetry permits only orthogonal response as in conventional SHE: the electric field, spin current, and spin polarization are mutually orthogonal; see Fig. 2(a). Consequently, only three independent spin-conductivity components are nonzero: $\sigma_{xy}^z$, $\sigma_{zx}^y$ and $\sigma_{yz}^x$. Herein lies the physical origin of the transport ambiguity: each nonzero component inextricably comprises both a T-even and a T-odd contribution. The T-even components correspond to the conventional SHE found in nonmagnetic metals, while the T-odd components arise from the MSHE enabled by the breaking of combined PT and T$\tau$ symmetries [11]. As shown in Figs. 2(c)-2(e), the calculated conductivities exhibit specific relationships when exchanging the $x$ and $y$ indices: the T-even components (blue symbols) are opposite while the T-odd components (red symbols) are equal. Specifically, we find $\sigma_{xy}^{\mathrm{even},z} = -\sigma_{yx}^{\mathrm{even},z}$ and $\sigma_{xy}^{\mathrm{odd},z} = \sigma_{yx}^{\mathrm{odd},z}$ for the $xy$ components, and $\sigma_{zx}^{\mathrm{even},y} = -\sigma_{zy}^{\mathrm{even},x}$ and $\sigma_{zx}^{\mathrm{odd},y} = \sigma_{zy}^{\mathrm{odd},x}$ for the $zx$ and $zy$ component pairs [Fig. 2(d)]. Analogous relations also hold for the $xz$ and $yz$ pairs in Fig. 2(e). These identities are strictly enforced by the magnetic space group symmetry of the superlattice. Owing to the tetragonal symmetry, the spin Hall conductivity is strongly anisotropic in magnitude, consistent with previous reports [39,60-62].

Remarkably, among all components, the T-odd conductivities $\sigma_{xy}^{\mathrm{odd},z} = \sigma_{yx}^{\mathrm{odd},z}$ are exceptionally large, exceeding their T-even counterparts and all other tensor elements by more than two orders of magnitude [Fig. 2(c)]. To elucidate its origin, we performed calculations without SOC. In this limit, all T-even contributions vanish, while sizable T-odd conductivities persist [red open circles and plus symbols in Fig. 2(c)], demonstrating their non-relativistic origin. These surviving contributions are attributed to the SSE generated by altermagnetism. As schematically illustrated in Fig. 2(b), an electric field applied along [100] induces different Fermi velocities for spin-up and spin-down carriers, generating a longitudinal charge current accompanied by a transverse pure spin current polarized along the Néel vector ($z||[001]$). The close quantitative agreement between $\sigma_{xy}^{\mathrm{odd},z}$ calculated with SOC and the spin-splitting contribution $\sigma_{xy}^{\mathrm{SSE},z}$ obtained without SOC confirms that this dominant response is overwhelmingly governed by altermagnetic spin splitting, while SOC plays only a secondary role. By contrast, for the remaining tensor components shown in Figs. 2(d) and 2(e), the spin polarization is orthogonal to the Néel vector, prohibiting contributions from the SSE mechanism and resulting in conductivities that are two orders of magnitude smaller. Consequently, the giant altermagnetic signature in this conventional geometry is entirely concentrated in the $xy$ components. However, returning to the core challenge of transport ambiguity, even this overwhelmingly large SSE signal intrinsically coexists with the T-even SHE background along the principal axes. This unavoidable physical entanglement necessitates a new symmetry-based strategy for absolute isolation.

*A symmetry-based transport criterion in low-symmetry configurations.* —To overcome the fundamental entanglement inherent in the principal axes, we must search for a symmetry-protected zero-background window by reducing the crystal symmetry, via fabricating single-crystalline films with high-Miller-index surface orientations and/or applying the electrical current along low-symmetry in-plane directions. This deliberate symmetry breaking unlocks unconventional transport phenomena forbidden in the high-symmetry geometry: a collinear spin current whose spin polarization is parallel to its propagating direction [Fig. 3(a)] (a key requirement for field-free switching of perpendicular magnetization [63,64]), and a spin-polarized electrical current [Fig. 3(b)] (despite the fully compensated magnetic structure).

We consider three representative configurations that host such unconventional spin currents: (i) a $(0\bar{1}1)$ film with $x||[100]$, (ii) a (101) film with $x||[10\bar{1}]$, and (iii) a (001) film with $x||[110]$. Each low-symmetry configuration can be generated by rotating the principal-axis system about one of its Cartesian axes, thereby preserving the symmetry along that axis while breaking the others. The corresponding spin-conductivity tensors are obtained by rotating the three independent nonzero components of the principal-axis tensor using appropriate rotation matrices [39], as shown in Fig. 3(c) and 3(d).

Crucially, a comparative analysis of these orientations reveals the precise origin of the ambiguity in previous experimental studies. In configurations (i) and (ii), including the widely investigated (101) plane, the magnetic symmetries do not involve time-reversal. Consequently, T-odd and T-even components are subject to identical symmetry constraints and thus coexist with the same angular dependencies. This shared symmetry leads to the inextricably mixed signals [blue and red bars in Figs. 3(c) and 3(d)] that have stalled the field. In striking contrast, configuration (iii) represents a conceptual breakthrough for experimental verification, because this geometry satisfies additional symmetry operations, $\left\{M_xT\middle|\frac{a}{2},\frac{b}{2},0\right\}$ and $\left\{M_yT\middle|\frac{a}{2},\frac{b}{2},0\right\}$ ($M_x$ and $M_y$ denote mirror operations about the $x$=0 and $y$=0 planes, respectively). These symmetries impose distinct constraints on the T-odd and T-even spin-conductivity tensors. The components $\sigma_{xz}^x$, $\sigma_{yz}^y$, $\sigma_{xx}^z$, and $\sigma_{yy}^z$ undergo a sign reversal under $M_x$ or $M_y$. Therefore, to remain invariant under the combined symmetries $M_xT$ and $M_yT$, these components must acquire an additional sign change under time reversal. Consequently, only the T-odd parts of $\sigma_{xz}^x$, $\sigma_{yz}^y$, $\sigma_{xx}^z$, and $\sigma_{yy}^z$ are symmetry allowed, whereas the corresponding T-even SHE backgrounds are strictly forced to zero. This rigorous symmetry analysis explains why exclusively T-odd components survive in configuration (iii), as explicitly demonstrated by the absence of blue bars in Figs. 3(c) and 3(d).

*A definitive roadmap for transport verification.*—The selective suppression of T-even contributions while retaining T-odd responses provides a symmetry-based transport criterion for identifying $d$-wave altermagnetism in $RuO_2$-type systems. This distinction arises because time-reversal symmetry is broken by magnetic order but preserved in nonmagnetic systems. Consequently, when crystal symmetry forbids T-even spin-transport contributions, the corresponding T-odd responses are also forbidden in a nonmagnetic material. In a magnetic system, by contrast, a finite symmetry-allowed T-odd response can therefore provide a direct transport signature of magnetic order. Furthermore, the same transport geometry enables the isolation of altermagnetic responses from potential parasitic signals associated with uncompensated interfacial moments [42]. This criterion directly addresses the central ambiguity in previous studies [18,33,39], by avoiding the need to infer the altermagnetic contribution from a symmetry-allowed background.

To translate the symmetry-based criterion into an experimentally accessible measurement, we focus on the longitudinal spin-polarized current in configuration (iii), which is more directly accessible experimentally than a collinear pure spin current. A possible implementation is a spin valve or magnetic tunnel junction (MTJ) comprising the $RuO_2$/$MnO_2$ superlattice and a ferromagnetic electrode separated by a nonmagnetic spacer or tunnel barrier. Reversing the magnetization of the ferromagnetic electrode with an external field would probe the spin polarization of the current through the resulting magnetoresistance. As a complementary control, the longitudinal spin-polarized currents can be measured for two orthogonal in-plane current directions, e.g. $[110]$ and $[\bar{1}10]$, and combined to extract their antisymmetric component. The symmetry analysis in the Supplemental Material [42] shows that contributions from interfacial uncompensated moments collinear with the Néel order are symmetric under this interchange and therefore cancel in the antisymmetric combination, whereas the *d*-wave altermagnetic response is antisymmetric. This provides an internal control for distinguishing the altermagnetic contribution from residual interfacial magnetic artifacts. Complementing this artifact control, we further examined the robustness of the target T-odd spin conductivity against representative materials imperfections, including interfacial chemical disorder and oxygen vacancies [42]. Our additional calculations show that the target T-odd response remains largely unchanged under these perturbations, further supporting the viability of the proposed $RuO_2/MnO_2$ superlattice as a platform for experimental verification.

To quantify this effect, we calculated the spin polarization of the ballistic current along [110] in $RuO_2/MnO_2$ superlattice. Within the scattering transport formalism [65,66], the spin-resolved Sharvin conductance along a particular direction $\hat{n} = [110]$ is given by

$$G_\sigma(\hat{n}) = \frac{e^2}{2\hbar N_k V} \sum_{m,\boldsymbol{k}} \left(-\frac{\partial f}{\partial \varepsilon}\right) |\hat{n} \cdot \nabla_{\boldsymbol{k}} \varepsilon_{m\boldsymbol{k}}^{\sigma}| \tag{3}$$

where $\sigma$ labels the spin channel. For $RuO_2/MnO_2$ superlattice, we obtain $G_\uparrow = 0.169 \times 10^{15}\ \Omega^{-1}\mathrm{m}^{-2}$ and $G_\downarrow = 0.083 \times 10^{15}\ \Omega^{-1}\mathrm{m}^{-2}$, corresponding to a current spin polarization $P_{\mathrm{AM}} = \frac{G_\uparrow - G_\downarrow}{G_\uparrow + G_\downarrow} = 34.1\%$.

We estimate the resulting TMR ratio using the Jullière model [67] $\mathrm{TMR} = \frac{2P_{\mathrm{AM}}P_{\mathrm{FM}}}{1-P_{\mathrm{AM}}P_{\mathrm{FM}}}$, where the spin polarizations $P_{\mathrm{FM}}$ of the common ferromagnetic electrodes are taken from experiment [68]. The calculated TMR ratios, summarized in Table I, range from approximately 29% for NiFe to nearly 90% for $CrO_2$. These sizeable values conclusively demonstrate that the symmetry-protected altermagnetic signals can be robustly and efficiently detected through standard magnetotransport measurements, offering a definitive and highly accessible roadmap for experimentalists.

*Conclusions.*— In summary, we have presented a comprehensive, two-pronged strategy to overcome the fundamental bottlenecks in stabilizing and verifying *d*-wave altermagnetism in $RuO_2$-based systems. To resolve the critical entanglement between spin-splitting effects and anisotropic spin Hall backgrounds, we established a rigorous, symmetry-based transport criterion. By identifying specific crystallographic geometries where T-even contributions are strictly forbidden, we provide a zero-background experimental window for the unambiguous isolation of pure altermagnetic responses. Concurrently, to bypass the highly controversial and fragile energetics of bulk $RuO_2$, we demonstrated superlattice engineering as a robust materials-design strategy. Using $RuO_2$/$MnO_2$ superlattices as a prototypical platform, we showed that interfacial exchange coupling rigidly enforces the altermagnetic order and pronounced spin splitting, independent of bulk uncertainties, and remains robust against variations in the superlattice period. Within this platform, our systematic spin-transport analysis revealed unconventional, orientation-dependent effects, including collinear spin currents capable of field-free switching of perpendicular magnetization and sizable spin-polarized electrical currents suitable for MTJ devices.

Ultimately, by combining a definitive, background-free transport criterion with a robustly engineered material platform, our work transcends the phenomenological debates over bulk energetics. It provides the community with an actionable, ambiguity-free roadmap for the experimental identification of *d*-wave altermagnetism, thereby breaking the current impasse surrounding $RuO_2$ and paving the way for reliable altermagnetic spintronics.

*Acknowledgements.*—This work was supported by the National Key Research and Development Program of China (2024YFA1408500), the National Natural Science Foundation of China (Grants No. 12574115 and No. 12374101), the Open Fund of the State Key Laboratory of Spintronics Devices and Technologies (Grant No. SPL-2408).

[1] L. Šmejkal, J. Sinova, and T. Jungwirth, Beyond conventional ferromagnetism and antiferromagnetism: A phase with nonrelativistic spin and crystal rotation symmetry, Phys. Rev. X **12**, 031042 (2022).
[2] L. Šmejkal, J. Sinova, and T. Jungwirth, Emerging research landscape of altermagnetism, Phys. Rev. X **12**, 040501 (2022).
[3] W. F. Brinkman and R. J. Elliott, Theory of spin-space groups, Proc. R. Soc. A. **294**, 343 (1966).
[4] D. B. Litvin, W. Opechowski, Spin groups, Physica **76**, 538 (1974).
[5] D. B. Litvin, Spin point groups, Acta Crystallogr. Sect. A: Found. Adv. **33**, 279 (1977).
[6] P. Liu, J. Li, J. Han, X. Wan, and Q. Liu, Spin-group symmetry in magnetic materials with negligible spin-orbit coupling, Phys. Rev. X **12**, 021016 (2022).
[7] Z. Xiao, J. Zhao, Y. Li, R. Shindou, and Z.-D. Song, Spin space groups: Full classification and applications, Phys. Rev. X **14**, 031037 (2024).
[8] X. Chen, J. Ren, Y. Zhu, Y. Yu, A. Zhang, P. Liu, J. Li, Y. Liu, C. Li, and Q. Liu, Enumeration and representation theory of spin space groups, Phys. Rev. X **14**, 031038 (2024).
[9] Y. Jiang, Z. Song, T. Zhu, Z. Fang, H. Weng, Z.-X. Liu, J. Yang, and C. Fang, Enumeration of spin-space groups: Toward a complete description of symmetries of magnetic orders, Phys. Rev. X **14**, 031039 (2024).
[10] J. Yang, Z.-X. Liu, and C. Fang, Symmetry invariants and classes of quasiparticles in magnetically ordered systems having weak spin-orbit coupling, Nat. Commun. **15**, 10203 (2024).
[11] R. González-Hernández, L. Šmejkal, K. Výborný, Y. Yahagi, J. Sinova, T. Jungwirth, and J. Železný, Efficient electrical spin splitter based on nonrelativistic collinear antiferromagnetism, Phys. Rev. Lett. **126**, 127701 (2021).
[12] S. Hayami, Y. Yanagi, and H. Kusunose, Momentum-dependent spin splitting by collinear antiferromagnetic ordering, J. Phys. Soc. Jpn. **88**, 123702 (2019).
[13] L.-D. Yuan, Z. Wang, J.-W. Luo, E. I. Rashba, and A. Zunger, Giant momentum-dependent spin splitting in centrosymmetric low-Z antiferromagnets, Phys. Rev. B **102**, 014422 (2020).
[14] H.-Y. Ma, M. Hu, N. Li, J. Liu, W. Yao, J.-F. Jia, and J. Liu, Multifunctional antiferromagnetic materials with giant piezomagnetism and noncollinear spin current, Nat. Commun. **12**, 2846 (2021).
[15] L. Bai, W. Feng, S. Liu, L. Šmejkal, Y. Mokrousov, and Y. Yao, Altermagnetism: Exploring new frontiers in magnetism and spintronics, Adv. Funct. Mater. **34**, 2409327 (2024).
[16] S. S. Fender, O. Gonzalez, and D. K. Bediako, Altermagnetism: A chemical perspective, J. Am. Chem. Soc. **147**, 2257 (2025).
[17] C. Song, H. Bai, Z. Zhou, L. Han, H. Reichlova, J. Hugo Dil, J. Liu, X. Chen, and F. Pan, Altermagnets as a new class of functional materials, Nat. Rev. Mater. **10**, 473 (2025).
[18] A. Bose, N. J. Schreiber, R. Jain, D.-F. Shao, H. P. Nair, J. Sun, X. S. Zhang, D. A. Muller, E. Y. Tsymbal, D. G. Schlom, and D. C. Ralph, Tilted spin current generated by the collinear antiferromagnet ruthenium dioxide, Nat. Electron. **5**, 267 (2022).
[19] M. Naka, S. Hayami, H. Kusunose, Y. Yanagi, Y. Motome, and H. Seo, Spin current generation in organic antiferromagnets, Nat. Commun. **10**, 4305 (2019).

[20] M. Naka, Y. Motome, and H. Seo, Perovskite as a spin current generator, Phys. Rev. B **103**, 125114 (2021).
[21] T. Berlijn, P. C. Snijders, O. Delaire, H.-D. Zhou, T. A. Maier, H.-B. Cao, S.-X. Chi, M. Matsuda, Y. Wang, M. R. Koehler, P. R. C. Kent, and H. H. Weitering, Itinerant antiferromagnetism in $RuO_2$, Phys. Rev. Lett. **118**, 077201 (2017).
[22] Z. H. Zhu, J. Strempfer, R. R. Rao, C. A. Occhialini, J. Pelliciari, Y. Choi, T. Kawaguchi, H. You, J. F. Mitchell, Y. Shao-Horn, and R. Comin, Anomalous antiferromagnetism in metallic $RuO_2$ determined by resonant x-ray scattering, Phys. Rev. Lett. **122**, 017202 (2019).
[23] Z. Lin, D. Chen, W. Lu, X. Liang, S. Feng, K. Yamagami, J. Osiecki, M. Leandersson, B. Thiagarajan, J. Liu, C. Felser, and J. Ma, Observation of giant spin splitting and d-wave spin texture in room temperature altermagnet $RuO_2$, arXiv:2402.04995 (2024).
[24] O. Fedchenko, J. Minár, A. Akashdeep, S. W. D'Souza, D. Vasilyev, O. Tkach, L. Odenbreit, Q. Nguyen, D. Kutnyakhov, N. Wind, L. Wenthaus, M. Scholz, K. Rossnagel, M. Hoesch, M. Aeschlimann, B. Stadtmüller, M. Kläui, G. Schönhense, T. Jungwirth, A. B. Hellenes, G. Jakob, L. Šmejkal, J. Sinova, and H.-J. Elmers, Observation of time-reversal symmetry breaking in the band structure of altermagnetic $RuO_2$, Sci. Adv. **10**, eadj4883 (2024).
[25] Z. Feng, X. Zhou, L. Šmejkal, L. Wu, Z. Zhu, H. Guo, R. González-Hernández, X. Wang, H. Yan, P. Qin, X. Zhang, H. Wu, H. Chen, Z. Meng, L. Liu, Z. Xia, J. Sinova, T. Jungwirth, and Z. Liu, An anomalous Hall effect in altermagnetic ruthenium dioxide, Nat. Electron. **5**, 735 (2022).
[26] L. Šmejkal, R. González-Hernández, T. Jungwirth, and J. Sinova, Crystal time-reversal symmetry breaking and spontaneous Hall effect in collinear antiferromagnets, Sci. Adv. **6**, eaaz8809 (2020).
[27] H. Bai, L. Han, X. Feng, Y. Zhou, R. Su, Q. Wang, L. Liao, W. Zhu, X. Chen, F. Pan, X. Fan, and C. Song, Observation of spin splitting torque in a collinear antiferromagnet $RuO_2$, Phys. Rev. Lett. **128**, 197202 (2022).
[28] S. Karube, T. Tanaka, D. Sugawara, N. Kadoguchi, M. Kohda, and J. Nitta, Observation of spin-splitter torque in collinear antiferromagnetic $RuO_2$, Phys. Rev. Lett. **129**, 137201 (2022).
[29] Y. Zhang, H. Bai, L. Han, C. Chen, Y. Zhou, C. H. Back, F. Pan, Y. Wang, and C. Song, Simultaneous high charge-spin conversion efficiency and large spin diffusion length in altermagnetic $RuO_2$, Adv. Funct. Mater. **34**, 2313332 (2024).
[30] Y. Guo, J. Zhang, Z. Zhu, Y.-Y. Jiang, L. Jiang, C. Wu, J. Dong, X. Xu, W. He, B. He, Z. Huang, L. Du, G. Zhang, K. Wu, X. Han, D.-F. Shao, G. Yu, and H. Wu, Direct and inverse spin splitting effects in altermagnetic $RuO_2$, Adv. Sci. **11**, 2400967 (2024).
[31] H. Bai, Y. C. Zhang, Y. J. Zhou, P. Chen, C. H. Wan, L. Han, W. X. Zhu, S. X. Liang, Y. C. Su, X. F. Han, F. Pan, and C. Song, Efficient spin-to-charge conversion via altermagnetic spin splitting effect in antiferromagnet $RuO_2$, Phys. Rev. Lett. **130**, 216701 (2023).
[32] Y. Liu, H. Bai, Y. Song, Z. Ji, S. Lou, Z. Zhang, C. Song, Q. Jin, Inverse altermagnetic spin splitting effect-induced terahertz emission in $RuO_2$, Adv. Opt. Mater. **11**, 2300177 (2023).

[33] C. T. Liao, Y. C. Wang, Y. C. Tien, S. Y. Huang, and D. R. Qu, Separation of inverse altermagnetic spin-splitting effect from inverse spin Hall effect in $RuO_2$, Phys. Rev. Lett. **133**, 056701 (2024).
[34] M. Hiraishi, H. Okabe, A. Koda, R. Kadono, T. Muroi, D. Hirai, and Z. Hiroi, Nonmagnetic ground state in $RuO_2$ revealed by muon spin rotation, Phys. Rev. Lett. **132**, 166702 (2024).
[35] P. Keßler, L. Garcia-Gassull, A. Suter, T. Prokscha, Z. Salman, D. Khalyavin, P. Manuel, F. Orlandi, I. I. Mazin, R. Valentí, and S. Moser, Absence of magnetic order in $RuO_2$: insights from μSR spectroscopy and neutron diffraction, npj Spintronics **2**, 50 (2024).
[36] J. Liu, J. Zhan, T. Li, J. Liu, S. Cheng, Y. Shi, L. Deng, M. Zhang, C. Li, J. Ding, Q. Jiang, M. Ye, Z. Liu, Z. Jiang, S. Wang, Q. Li, Y. Xie, Y. Wang, S. Qiao, J. Wen, Y. Sun, and D. Shen, Absence of altermagnetic spin splitting character in rutile oxide $RuO_2$, Phys. Rev. Lett. **133**, 176401 (2024).
[37] M. Wenzel, E. Uykur, S. Rößler, M. Schmidt, O. Janson, A. Tiwari, M. Dressel, and A. A. Tsirlin, Fermi-liquid behavior of nonaltermagnetic $RuO_2$, Phys. Rev. B **111**, L041115 (2025).
[38] Z. Wu, M. Long, H. Chen, S. Paul, H. Matsuki, O. Zheliuk, U. Zeitler, G. Li, R. Zhou, Z. Zhu, D. Graf, T. I. Weinberger, F. M. Grosche, Y. Maeno, and A. G. Eaton, Fermi surface of $RuO_2$ measured by quantum oscillations, Phys. Rev. X, **15**, 031044 (2025).
[39] Z. Q. Wang, Z. Q. Li, L. Sun, Z. Y. Zhang, K. He, H. Niu, J. Cheng, M. Yang, X. Yang, G. Chen, Z. Yuan, H. F. Ding, and B. F. Miao, Inverse spin Hall effect dominated spin-charge conversion in (101) and (110)-oriented $RuO_2$ films, Phys. Rev. Lett. **133**, 046701 (2024).
[40] Y.-C. Wang, Z.-Y. Shen, C.-H. Lin, W.-C. Hsu, Y.-S. Chen, Y.-Y. Chin, A. K. Singh, W.-L. Lee, C.-T. Chen, S.-Y. Huang, and D. Qu, Absence of transport altermagnetic spin-splitting effect in $RuO_2$, Nano Lett. **26**, 2548-2554 (2026).
[41] K. Shibuya, M. Kawasaki, and Y. Tokura, Metal-insulator transitions in $TiO_2/VO_2$ superlattices, Phys. Rev. B **82**, 205118 (2010).
[42] See Supplemental Material at [URL] for computational details, superlattice size convergence, magnetic and structural stability, symmetry elements and allowed spin conductivity responses, as well as other technical issues, which includes Refs. [43-51].
[43] G. Kresse and J. Furthmüller, Efficient iterative schemes for ab initio total-energy calculations using a plane-wave basis set, Phys. Rev. B **54**, 11169 (1996).
[44] G. Kresse and D. Joubert, From ultrasoft pseudopotentials to the projector augmented-wave method, Phys. Rev. B **59**, 1758 (1999).
[45] P. E. Blöchl, Projector augmented-wave method, Phys. Rev. B **50**, 17953 (1994).
[46] J. P. Perdew, K. Burke, and M. Ernzerhof, Generalized gradient approximation made simple, Phys. Rev. Lett. **77**, 3865 (1996).
[47] S. L. Dudarev, G. A. Botton, S. Y. Savrasov, C. J. Humphreys, and A. P. Sutton, Electron-energy-loss spectra and the structural stability of nickel oxide: An LSDA+U study, Phys. Rev. B **57**, 1505 (1998).
[48] Y. Noda, K. Ohno, and S. Nakamura, Momentum-dependent band spin splitting in semiconducting MnO2: a density functional calculation, Phys. Chem. Chem. Phys. **18**, 13294 (2016).
[49] M. Chepkoech, D. P. Joubert, and G. O. Amolo, First principles calculations of the

thermoelectric properties of $\alpha$-$MnO_2$ and $\beta$-$MnO_2$, Eur. Phys. J. B **91**, 301 (2018).
[50] A. Gauswami and P. K. Jha, Strain-controlled chiral magnons and spin-dependent Seebeck effect in tetragonal $\beta$-$MnO_2$ altermagnet: a DFT study, J. Phys.: Condens. Matter **38**, 025802 (2026).
[51] A. A. Mostofi, J. R. Yates, G. Pizzi, Y.-S. Lee, I. Souza, D. Vanderbilt, and N. Marzari, An updated version of wannier90: A tool for obtaining maximally-localised Wannier functions, Comput. Phys. Commun. **185**, 2309 (2014).
[52] A. Smolyanyuk, I. I. Mazin, L. Garcia-Gassull, and R. Valentí, Fragility of the magnetic order in the prototypical altermagnet $RuO_2$, Phys. Rev. B **109**, 134424 (2024).
[53] Z. Qian, Y. Yang, S. Liu, and C. Wu, Fragile unconventional magnetism in $RuO_2$ by proximity to Landau-Pomeranchuk instability, Phys. Rev. B **111**, 174425 (2025).
[54] Z. Zhu, R. Huang, X. Chen, Z. Cui, X. Duan, J. Zhang, I. Žutić, and T. Zhou, Altermagnetic proximity effect, Phys. Rev. Lett. **136**, 186702 (2026).
[55] J. Qiao, J. Zhou, Z. Yuan, and W. Zhao, Calculation of intrinsic spin Hall conductivity by Wannier interpolation, Phys. Rev. B **98**, 214402 (2018).
[56] F. Freimuth, S. Blügel, and Y. Mokrousov, Spin-orbit torques in Co/Pt(111) and Mn/W(001) magnetic bilayers from first principles, Phys. Rev. B **90**, 174423 (2014).
[57] J. Železný, Y. Zhang, C. Felser, and B. Yan, Spin-polarized current in noncollinear antiferromagnets, Phys. Rev. Lett. **119**, 187204 (2017).
[58] M. Kimata, H. Chen, K. Kondou, S. Sugimoto, P. K. Muduli, M. Ikhlas, Y. Omori, T. Tomita, A. H. MacDonald, S. Nakatsuji, and Y. Otani, Magnetic and magnetic inverse spin Hall effects in a non-collinear antiferromagnet, Nature **565**, 627 (2019).
[59] A. Mook, R. R. Neumann, A. Johansson, J. Henk, and I. Mertig, Origin of the magnetic spin Hall effect: Spin current vorticity in the Fermi sea, Phys. Rev. Research **2**, 023065 (2020).
[60] E. M. Chudnovsky, Intrinsic spin Hall effect in noncubic crystals, Phys. Rev. B **80**, 153105 (2009).
[61] F. Freimuth, S. Blügel, and Y. Mokrousov, Anisotropic spin Hall effect from first principles, Phys. Rev. Lett. **105**, 246602 (2010).
[62] Y. Sun, Y. Zhang, C.-X. Liu, C. Felser, and B. Yan, Dirac nodal lines and induced spin Hall effect in metallic rutile oxides, Phys. Rev. B **95**, 235104 (2017).
[63] D. MacNeill, G. M. Stiehl, M. H. D. Guimaraes, R. A. Buhrman, J. Park, and D. C. Ralph, Control of spin-orbit torques through crystal symmetry in $WTe_2$/ferromagnet bilayers, Nat. Phys. **13**, 300 (2017).
[64] M. Yang, L. Sun, Y. Zeng, J. Cheng, K. He, X. Yang, Z. Wang, L. Yu, H. Niu, T. Ji, G. Chen, B. Miao, X. Wang, and H. Ding, Highly efficient field-free switching of perpendicular yttrium iron garnet with collinear spin current, Nat. Commun. **15**, 3201 (2024).
[65] I. I. Mazin, How to define and calculate the degree of spin polarization in ferromagnets, Phys. Rev. Lett. **83**, 1427 (1999).
[66] K. M. Schep, P. J. Kelly, and G. E. W. Bauer, Ballistic transport and electronic structure, Phys. Rev. B **57**, 8907 (1998).
[67] M. Julliere, Tunneling between ferromagnetic films, Phys. Lett. A **54**, 225 (1975).
[68] R. J. Soulen Jr., J. M. Byers, M. S. Osofsky, B. Nadgorny, T. Ambrose, S. F. Cheng, P. R.

Broussard, C. T. Tanaka, J. Nowak, J. S. Moodera, A. Barry, and J. M. D. Coey, Measuring the spin polarization of a metal with a superconducting point contact, Science **282**, 85 (1998).

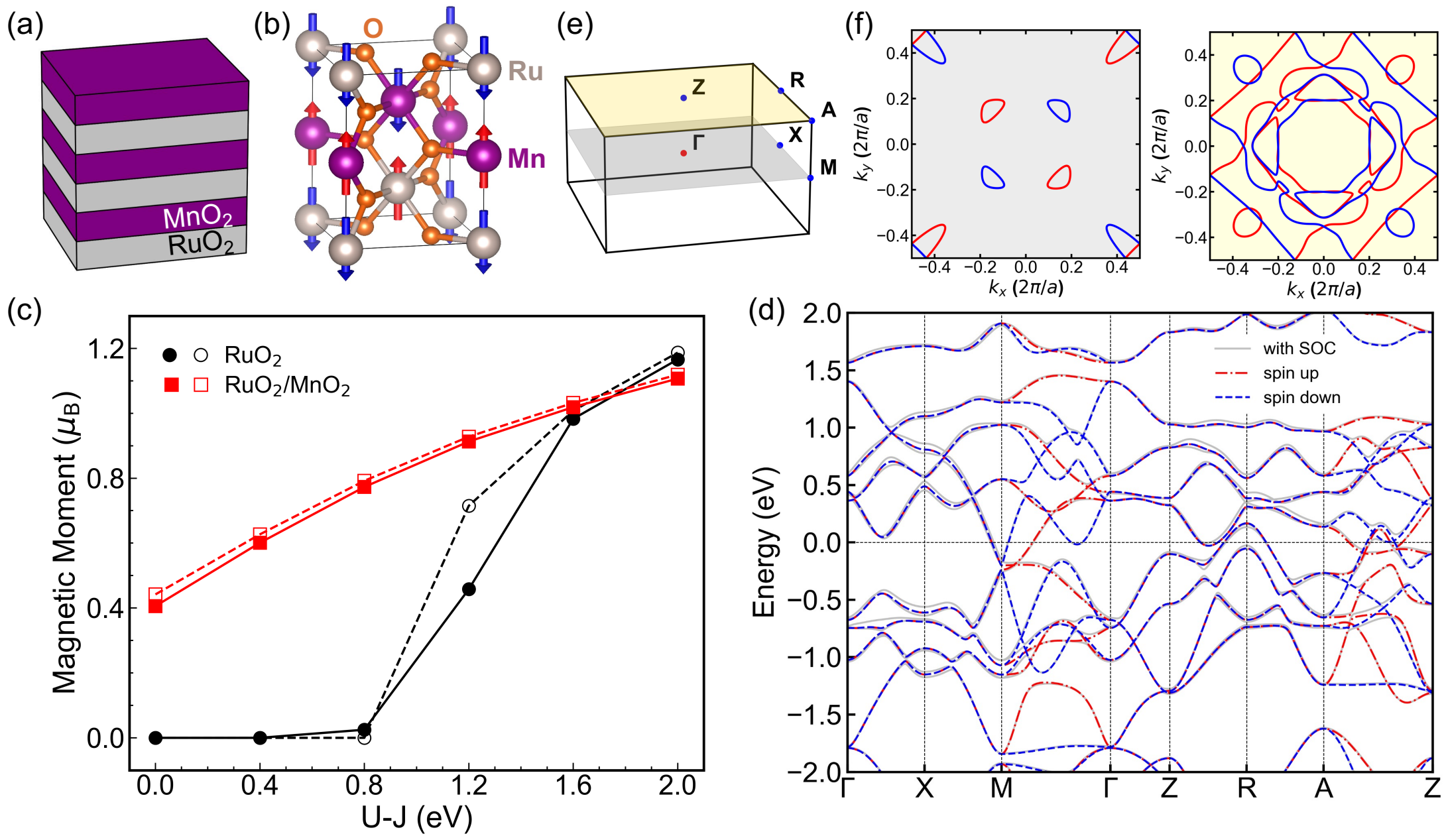

Fig. 1. (a) Schematic and (b) atomic structure of the $RuO_2/MnO_2$ superlattice. (c) Calculated magnetic moments on each Ru site vs effective Hubbard interaction $U - J$. Results for bulk $RuO_2$ (black circles) are plotted for comparison. Solid (empty) symbols denote calculations with (without) SOC. (d) Band structure of the $RuO_2/MnO_2$ superlattices calculated with $U - J = 0$ eV on Ru sites. Red and blue curves indicate spin-up and spin-down bands obtained without SOC; gray curves show the corresponding bands including SOC. (e) The first Brillouin zone of the $RuO_2/MnO_2$ superlattice with $k_z = 0$ (gray) and $k_z = \pi/c$ (yellow) planes. (f) Spin-resolved Fermi surface contours within the $k_z = 0$ (left) and $k_z = \pi/c$ (right) planes. Red and blue curves represent the Fermi surface contours for spin-up and spin-down electrons, respectively.

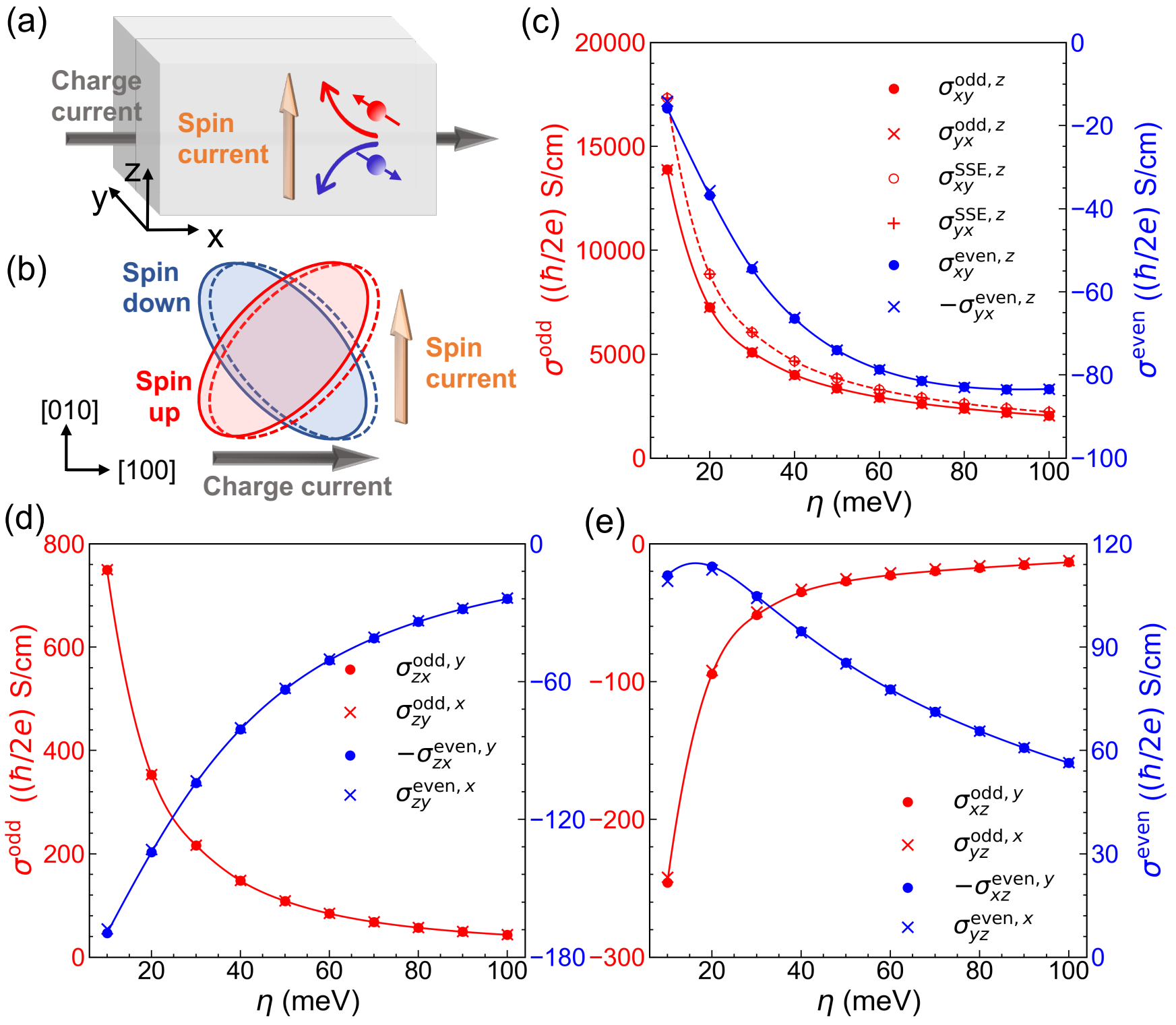

Fig. 2. (a) Schematic illustration of the conventional SHE. (b) Schematic of Fermi-surface asymmetry induced by altermagnetic spin splitting. (c-e) Calculated spin-conductivity components as a function of the broadening parameter $\eta$. Red (blue) symbols denote T-odd (T-even) contributions. SOC is included unless otherwise indicated; empty circles or plus symbols represent non-relativistic SSE contributions without SOC.

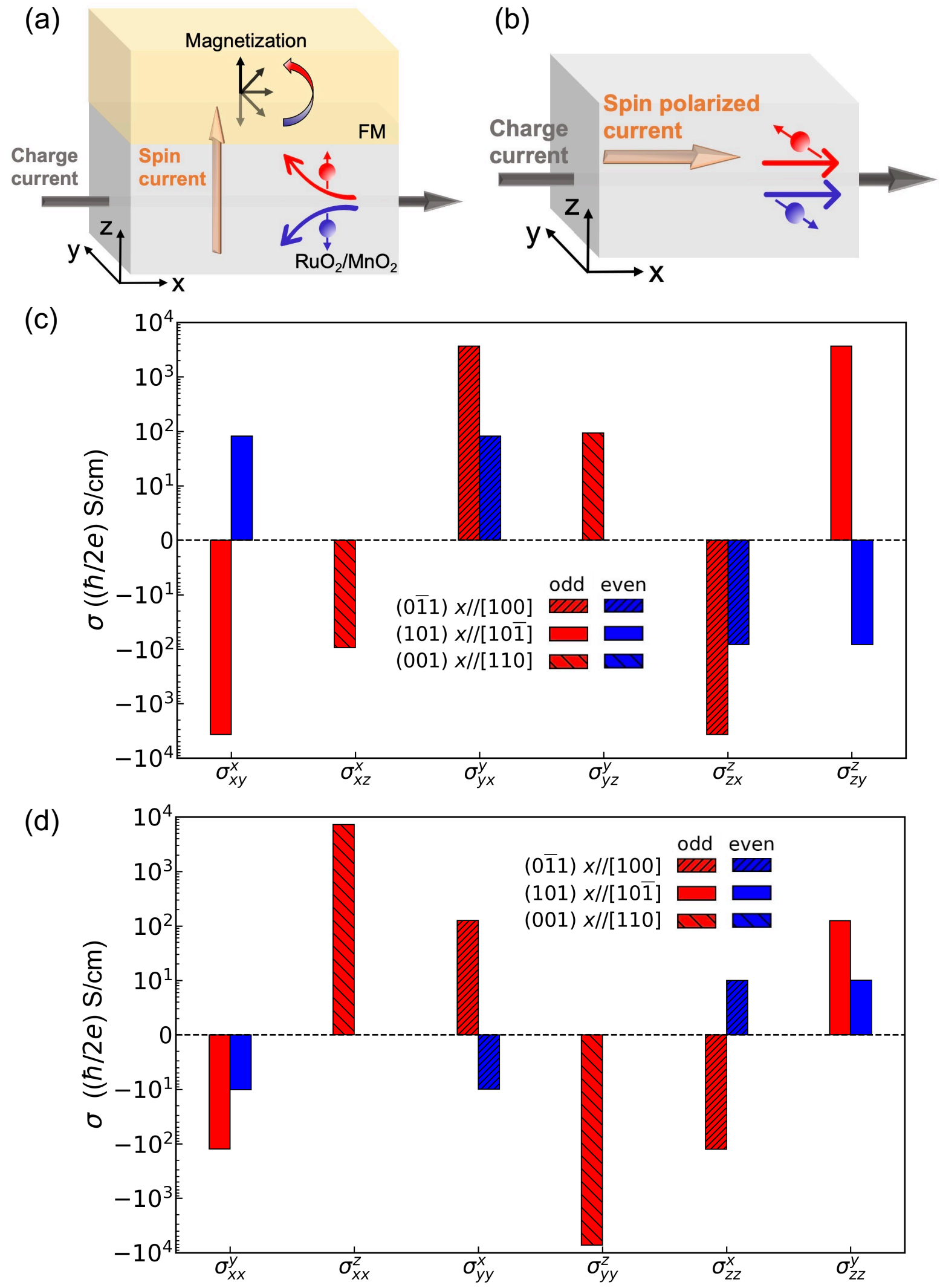

Fig. 3. Schematics of (a) collinear spin current and (b) spin-polarized charge current. Calculated (c) collinear spin-current components and (d) spin-polarized charge-current components for three representative crystallographic configurations. Red and blue bars denote T-odd and T-even contributions, respectively. Calculations were performed with a broadening parameter $\eta$ = 20 meV.

Table I. Estimated TMR ratios for magnetic tunnel junctions composed of a $RuO_2/MnO_2$ superlattice and various ferromagnetic electrodes. The spin polarizations of ferromagnetic electrodes are taken from Ref. [68].

| Ferromagnetic Electrode | Spin Polarization of Ferromagnetic Electrode (%) | TMR Ratio (%) |
|---|---|---|
| Fe | 45 | 36.3 |
| Co | 42 | 33.5 |
| Ni | 46.5 | 37.7 |
| NiFe | 37 | 28.9 |
| NiMnSb | 58 | 49.4 |
| LSMO | 78 | 72.6 |
| $CrO_2$ | 90 | 88.7 |

# Supplemental Material for "Symmetry-Based Transport Criterion and Robust Design for $RuO_2$-type Altermagnets"

Zhaoqing Li,[1] Yi Liu,[2,3,*] Yizheng Wu,[4] Zhe Yuan[1,†]

[1]*State Key Laboratory of Surface Physics and Interdisciplinary Center for Theoretical Physics and Information Sciences, Fudan University, Shanghai 200433, China*

[2]*Institute for Quantum Science and Technology, Shanghai University, Shanghai 200444, China*

[3]*Department of Physics, Shanghai University, Shanghai 200444, China*

[4]*State Key Laboratory of Surface Physics and Department of Physics, Fudan University, Shanghai 200433, China*

[*] yiliu42@shu.edu.cn; [†] yuanz@fudan.edu.cn

**A list of contents**

**Note 1. Computational details**

All self-consistent electronic structure calculations were performed within the framework of density functional theory (DFT) [1,2]. The projector augmented-wave (PAW) [3] method was employed to describe the electron-ion interactions, and the Perdew-Burke-Ernzerhof (PBE) functional within the generalized gradient approximation (GGA) was used for the exchange-correlation potential [4]. The valence states of Ru, Mn, and O were treated as $4d^7 5s^1$, $3d^6 4s^1$, and $2s^2 2p^4$, respectively. The plane-wave energy cutoff was set to 600 eV, and a 16×16×12 $k$-point mesh was used to sample the Brillouin zone. Prior to the static electronic structure calculations, both the atomic positions and the cell volumes of the superlattices were fully relaxed. The structural optimization was performed until the Hellmann-Feynman forces on each atom were less than 0.01 eV/Å and the total energy convergence criterion between self-consistent electronic steps was set to $10^{-7}$ eV. To account for electron correlation effects, the DFT+U method [5] was employed. The effective Hubbard parameter $U_{\mathrm{eff}} = U - J$ was fixed at 3.0 eV for the Mn $3d$ orbitals, which is consistent with previous studies on rutile $MnO_2$ [6-8] and successfully reproduces its semiconducting nature and local magnetic moments. For the Ru $4d$ orbitals, $U_{\mathrm{eff}}$ was varied from 0 to 2 eV following recent evaluations of bulk $RuO_2$ [9]. This systematic scan ensures that the altermagnetic order in the $RuO_2$/$MnO_2$ superlattice remains robust across the entire parameter regime where the magnetism of bulk $RuO_2$ is currently debated.

For the subsequent transport calculations, we applied $U_{\mathrm{eff}}$ = 3.0 eV only to Mn, with no additional $U$ correction on Ru ($U_{\mathrm{eff}}$ = 0 eV). Maximally localized Wannier functions (MLWFs) were constructed using WANNIER90 [10]. Both time-reversal-even (T-even) and time-reversal-odd (T-odd) components of the spin conductivity tensor were evaluated using Eqs. (1) and (2) in the main text. To ensure numerical convergence, the quantities were interpolated onto an ultra-dense 100 × 100 × 100 $k$-point mesh combined with a 3 × 3 × 3 adaptive refinement $k$ mesh. The convergence of the calculated $\sigma_{xy}^{\mathrm{odd},z}$ and $\sigma_{xy}^{\mathrm{even},z}$ with respect to the Wannier interpolation $k$-mesh density is demonstrated in Fig. S1, where the curves for meshes denser than 60 × 60 × 60 become indistinguishable, indicating robust convergence. Furthermore, the saturation of the spin conductivity at high mesh densities (the insets) confirms that both symmetry-distinct transport coefficients are well-converged with the 100 × 100 × 100 mesh used in the main text.

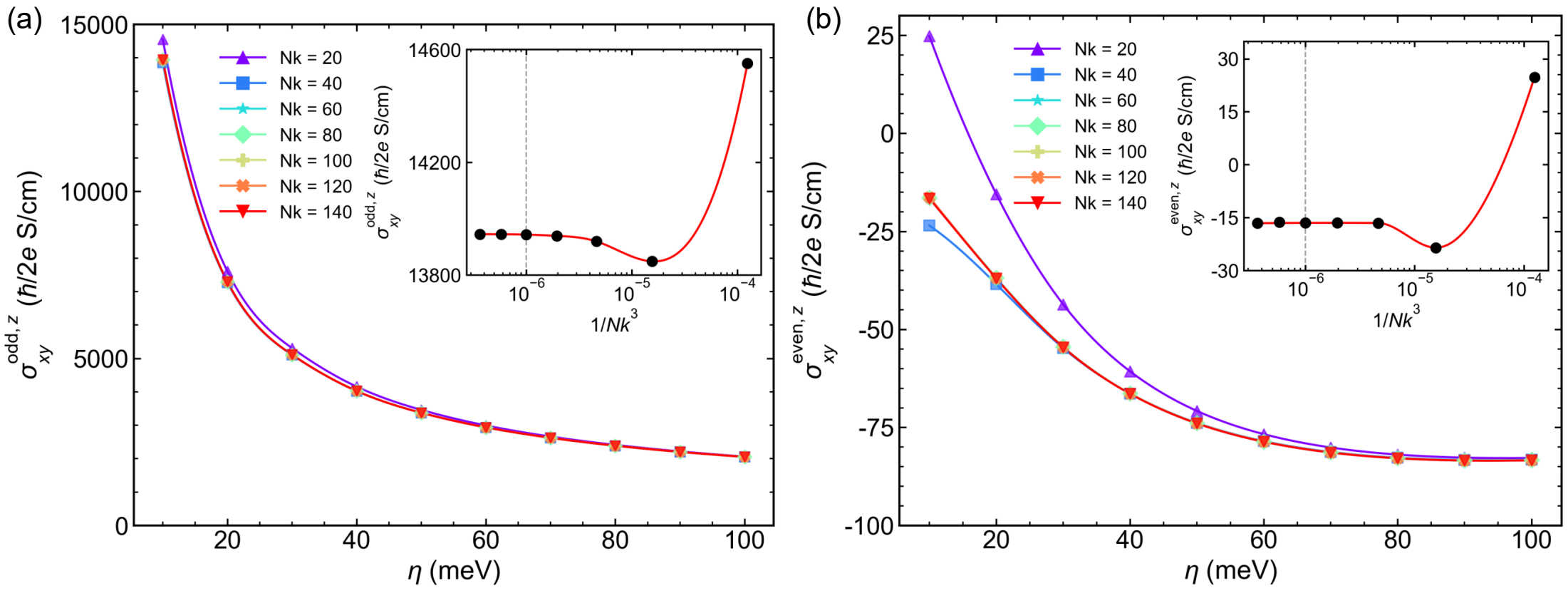

Fig. S1. Convergence tests of the spin conductivity with respect to the Wannier interpolation *k*-point mesh density. Calculated (a) T-odd spin Hall conductivity $\sigma_{xy}^{\text{odd},z}$ and (b) T-even spin Hall conductivity $\sigma_{xy}^{\text{even},z}$ as a function of the broadening parameter $\eta$ for various *k*-point meshes ($N_k \times N_k \times N_k$) ranging from 20 × 20 × 20 to 140 × 140 × 140. The insets display the corresponding conductivity values at a fixed broadening $\eta = 10\ \text{meV}$ plotted against $1/N_k^3$.

## Note 2. Superlattice period convergence

To verify that the predicted altermagnetic order is not an artifact of the specific superlattice period, we performed additional calculations by varying the layer thickness $n$ in the $(RuO_2)_n/(MnO_2)_n$ superlattices from $n$ = 1 to $n$ = 4. The calculations were performed without additional Hubbard U correction on Ru ($U_{\text{eff}}^{\text{Ru}}$= 0 eV) to test the robustness of the proximity-induced magnetism. Our results indicate that for all configurations ($n$=1 to 4), the system maintains a zero net magnetization, with the magnetic moments on the Ru and Mn sublattices remaining fully compensated. Crucially, while the magnitude of the induced local magnetic moments on Ru sites varies with the distance from the interface in thicker superlattices, the altermagnetic nature of the electronic structure is preserved. As shown in Fig. S2, pronounced spin splitting is consistently observed along the high-symmetry paths M–Γ and A–Z in the band structures calculated without spin-orbit coupling (SOC). This persistence of spin splitting, coexisting with zero net magnetization, confirms that the *d*-wave altermagnetism is a robust property of the $RuO_2/MnO_2$ superlattice system, independent of the layer thickness. Therefore, the $n$ = 1 configuration used in the main text is a representative model that captures the essential physics of the interface-stabilized altermagnetism while maintaining computational efficiency.

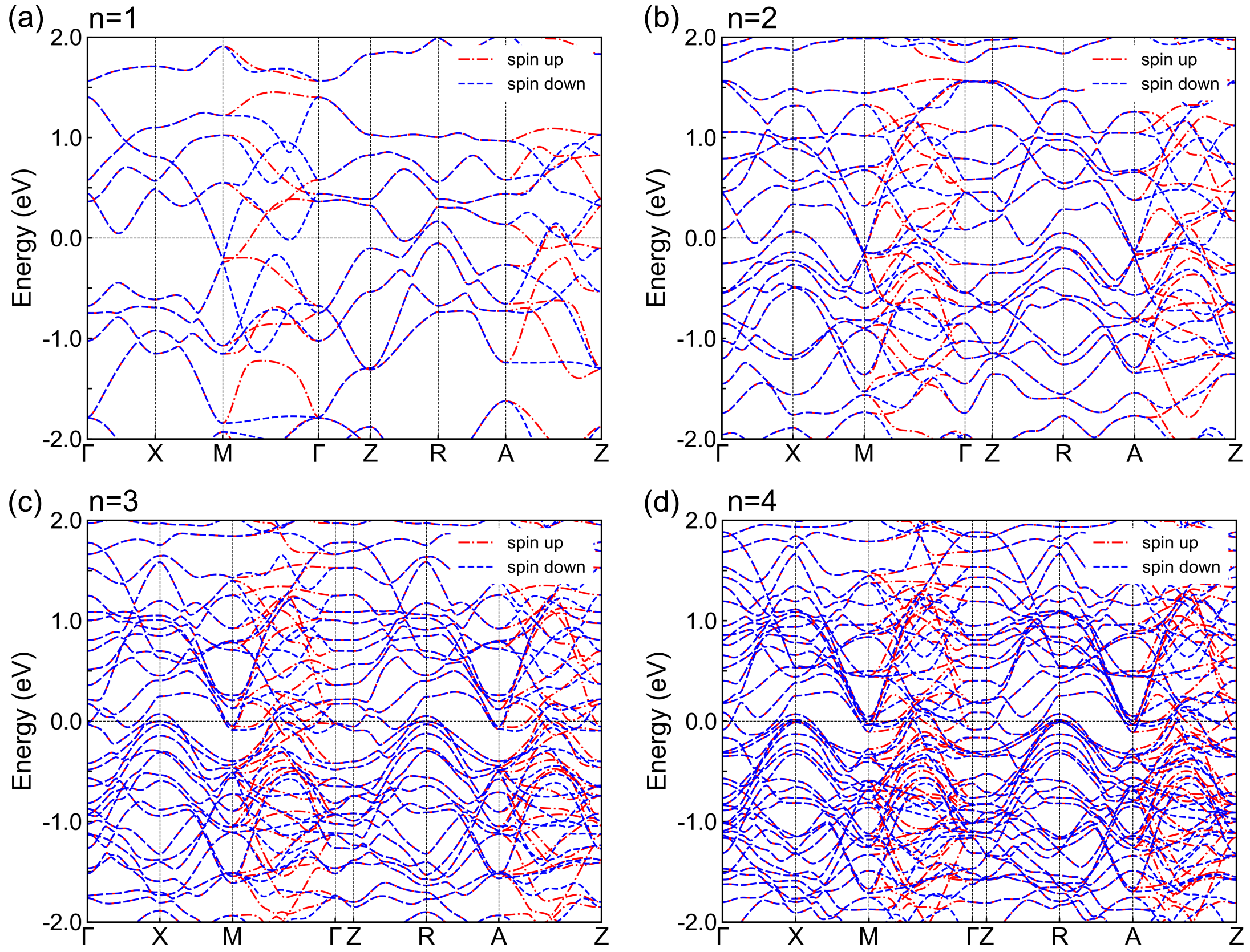

Fig. S2. Robustness of altermagnetic electronic structure against superlattice period. Calculated band structures of $(RuO_2)_n/(MnO_2)_n$ superlattices for (a) n=1, (b) n=2, (c) n=3, and (d) n=4. All calculations were performed without SOC and with $U_{\mathrm{eff}}^{\mathrm{Ru}}$ = 0 eV. The red and blue curves represent spin-up and spin-down bands, respectively.

**Note 3. Magnetic stability and sublattice magnetic moments**

To further assess the magnetic stability of the $(RuO_2)_n/(MnO_2)_n$ superlattices, we first compare the total energies of the altermagnetic (AM) and ferromagnetic (FM) states. As shown in Fig. S3(a), the energy difference $E_{\mathrm{FM}} - E_{\mathrm{AM}}$ for the $n$=1 superlattice remains positive and increases with the effective Hubbard parameter $U$-$J$ applied to the Ru $d$ orbitals. This indicates that the AM state is energetically favored over the FM state, with the energetic preference enhanced by the Ru-site correlation. Figure S3(b) further shows that $E_{\mathrm{FM}} - E_{\mathrm{AM}}$ increases with the superlattice period $n$ at $U$-$J$=0 eV on Ru sites, indicating that the energetic preference for the AM state persists and becomes more pronounced with increasing superlattice period.

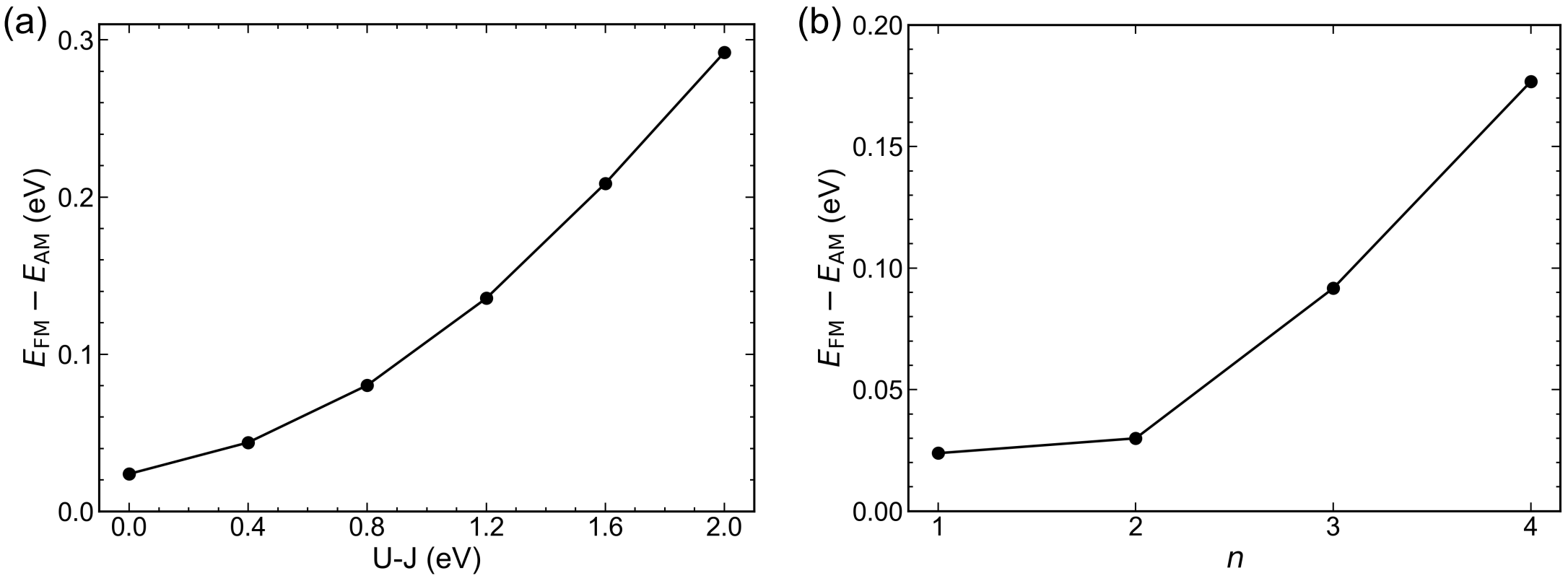

Fig. S3. Total energy difference between the FM and AM states, $E_{\mathrm{FM}} - E_{\mathrm{AM}}$, for $(RuO_2)_n/(MnO_2)_n$ superlattices. (a) Energy difference as a function of the effective Hubbard parameter $U$-$J$ on Ru sites for the $n$=1 superlattice. (b) Energy difference as a function of the superlattice period $n$ with $U$-$J$ = 0 eV on Ru sites. Positive energy differences throughout the considered parameter range indicate that the AM state remains energetically favored over the FM state.

Furthermore, analysis of the site-resolved local magnetic moments, summarized in Table S1, shows that the magnetic order is stabilized by the interfacial magnetic proximity effect. The $MnO_2$ layers retain antiferromagnetic order with large local moments. Through strong interfacial Ru-O-Mn superexchange, this magnetic template generates a staggered effective Zeeman field in the $RuO_2$ layers. The resulting proximity effect is reflected in the overall spatial decay of the induced Ru magnetic moments with increasing distance from the $MnO_2$ interface. This interfacial pinning anchors the Ru spins to the $MnO_2$ template and suppresses spontaneous magnetic reconstruction within the in-plane directions. Table S1 further shows that the magnetic moments of the Ru and Mn atoms compensate within each constituent layer, resulting in zero total magnetization for all investigated superlattice periods.

Table S1. Calculated local magnetic moments (in $\mu_B$) for Ru and Mn sites in $(RuO_2)_n/(MnO_2)_n$ superlattices with $n$ = 1–4 (at $U$-$J$ = 0 eV on Ru sites). The atomic sites (Ru1-Ru8, Mn1-Mn8) are strictly indexed along the out-of-plane $z$-axis.

| | n=1 | n=2 | n=3 | n=4 |
|---|---|---|---|---|
| $M_{Ru1}$ | −0.442 | +0.300 | +0.208 | +0.185 |
| $M_{Ru2}$ | +0.442 | −0.158 | −0.050 | −0.012 |
| $M_{Ru3}$ | - | +0.158 | +0.058 | +0.009 |
| $M_{Ru4}$ | - | −0.300 | −0.058 | +0.077 |
| $M_{Ru5}$ | - | - | +0.050 | −0.077 |
| $M_{Ru6}$ | - | - | −0.208 | −0.009 |
| $M_{Ru7}$ | - | - | - | +0.012 |
| $M_{Ru8}$ | - | - | - | −0.185 |
| $M_{Mn1}$ | +3.059 | +3.039 | +3.043 | +3.041 |
| $M_{Mn2}$ | −3.059 | −2.883 | −2.893 | −2.891 |
| $M_{Mn3}$ | - | +2.883 | +2.870 | +2.868 |
| $M_{Mn4}$ | - | −3.039 | −2.870 | −2.875 |
| $M_{Mn5}$ | - | - | +2.893 | +2.875 |
| $M_{Mn6}$ | - | - | −3.043 | −2.868 |
| $M_{Mn7}$ | - | - | - | +2.891 |
| $M_{Mn8}$ | - | - | - | −3.041 |
| $M_{total}$ | 0 | 0 | 0 | 0 |

To further assess the stability of the AM state against alternative antiferromagnetic (AFM) orderings, we compare its total energy with those of four representative competing AFM states for $n$ = 1, 2, 3, and 4. As illustrated in Fig. S4(a)–(b), these AFM states (AFM1–AFM4) were constructed within an in-plane $2 \times 2$ supercell, with the Mn magnetic moments arranged according to the corresponding spin pattern and the induced Ru moments relaxed self-consistently. Figure S4(c) summarizes the calculated total-energy differences $E_{\mathrm{AFM}} - E_{\mathrm{AM}}$ per $(RuO_2)_n/(MnO_2)_n$ superlattice unit cell. For all four representative AFM states, $E_{\mathrm{AFM}} - E_{\mathrm{AM}}$ remains positive for $n$ = 1–4, indicating that the AM state is energetically favored over each of these competing states. However, the minimum energy difference for $n$ = 1 is only approximately 7 meV per superlattice unit cell, suggesting that the AM state in this shortest-period superlattice may be susceptible to thermal effects. For larger periods, the minimum energy differences increases, indicating a progressively stronger energetic preference for the AM state. These results, together with the AM–FM energy comparisons above, support the stability of the AM state across the investigated superlattice periods.

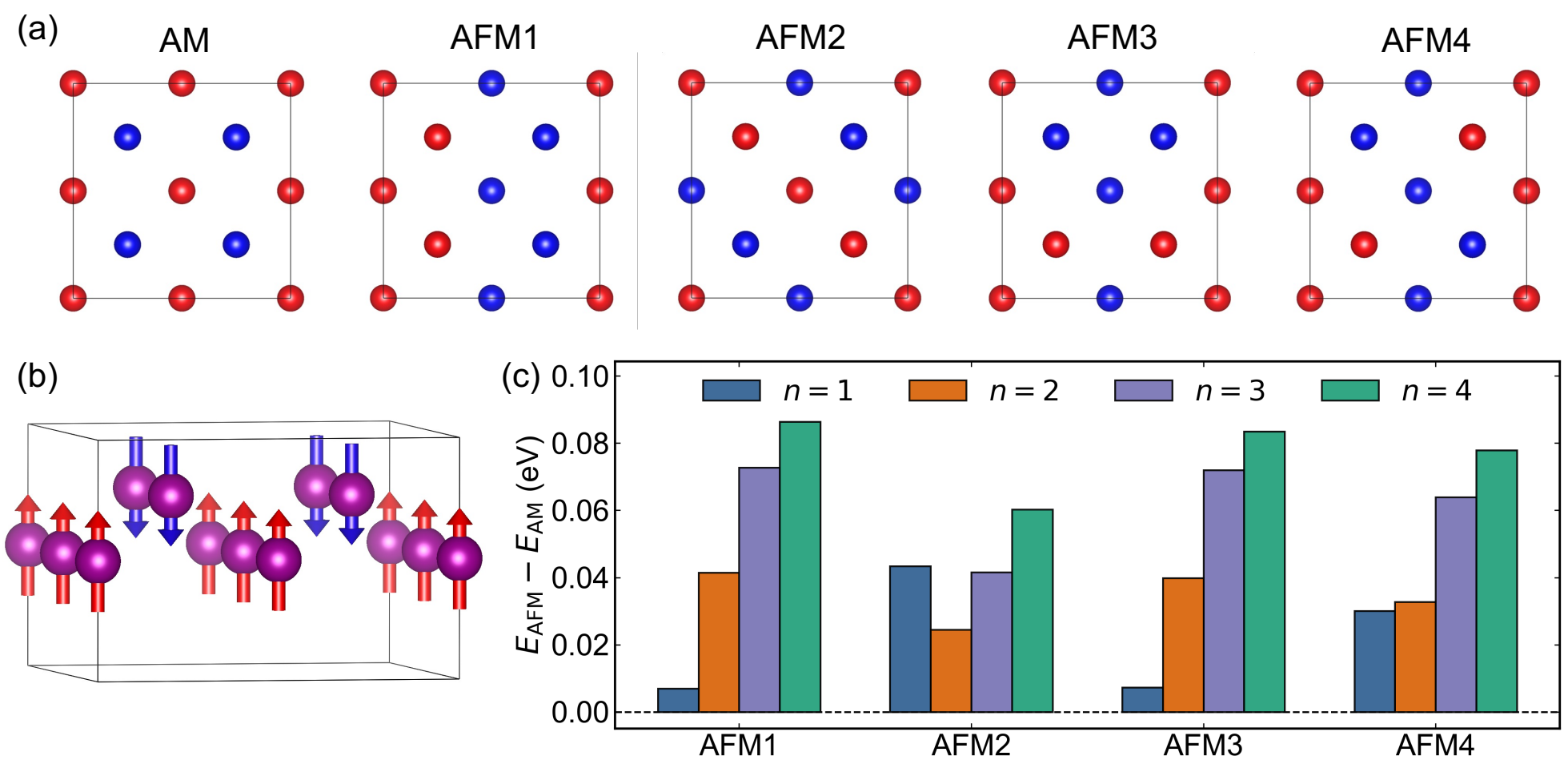

Fig. S4. (a) Schematic top views of the Mn magnetic-moment arrangements within one repeating block for the AM state and four representative competing antiferromagnetic states (AFM1–AFM4). Red and blue spheres represent Mn atoms with spin-up and spin-down magnetic moments, respectively. (b) Perspective view of the Mn sublattices for the AM state, with red and blue arrows indicating the directions of the local magnetic moments. (c) Calculated total-energy differences $E_{\mathrm{AFM}} - E_{\mathrm{AM}}$ per $(RuO_2)_n/(MnO_2)_n$ superlattice unit cell for $n$ = 1, 2, 3, and 4. The small energy differences for $n$ = 1 highlight the sensitivity of this shortest-period superlattice to finite-temperature effects.

## Note 4. Thermodynamic and dynamical stability of $(RuO_2)_n/(MnO_2)_n$ superlattices

To comprehensively verify the physical stability of the proposed $(RuO_2)_n/(MnO_2)_n$ superlattices, we investigated both their thermodynamic and dynamical properties using first-principles calculations.

First, to address the thermodynamic competitiveness of the superlattice against phase segregation into separate oxides, we calculated the formation energy defined as $E_{formation} = E_{superlattice} - (E_{bulk_RuO2} + E_{bulk_MnO2})$. Our calculations yield a negative formation energy of approximately -10.11 meV/atom. This intrinsic energetic driving force indicates that the periodic superlattice is stable , effectively preventing spontaneous phase separation.

Furthermore, we performed comprehensive phonon dispersion calculations to verify the dynamical stability. As shown in Fig. S5, for all investigated superlattice periods ($n$ = 1, 2, 3, 4), the phonon branches reside entirely within the positive frequency domain throughout the Brillouin zone. The complete absence of imaginary (soft) modes definitively proves the dynamical stability of the $(RuO_2)_n/(MnO_2)_n$ system. This robust stability is consistent with the identical rutile symmetry and the small lattice mismatch (≈1.8%) between the constituent layers. These findings suggest that the superlattices are experimentally realizable via modern epitaxial techniques and provide a reliable platform for investigating altermagnetic spin-transport phenomena.

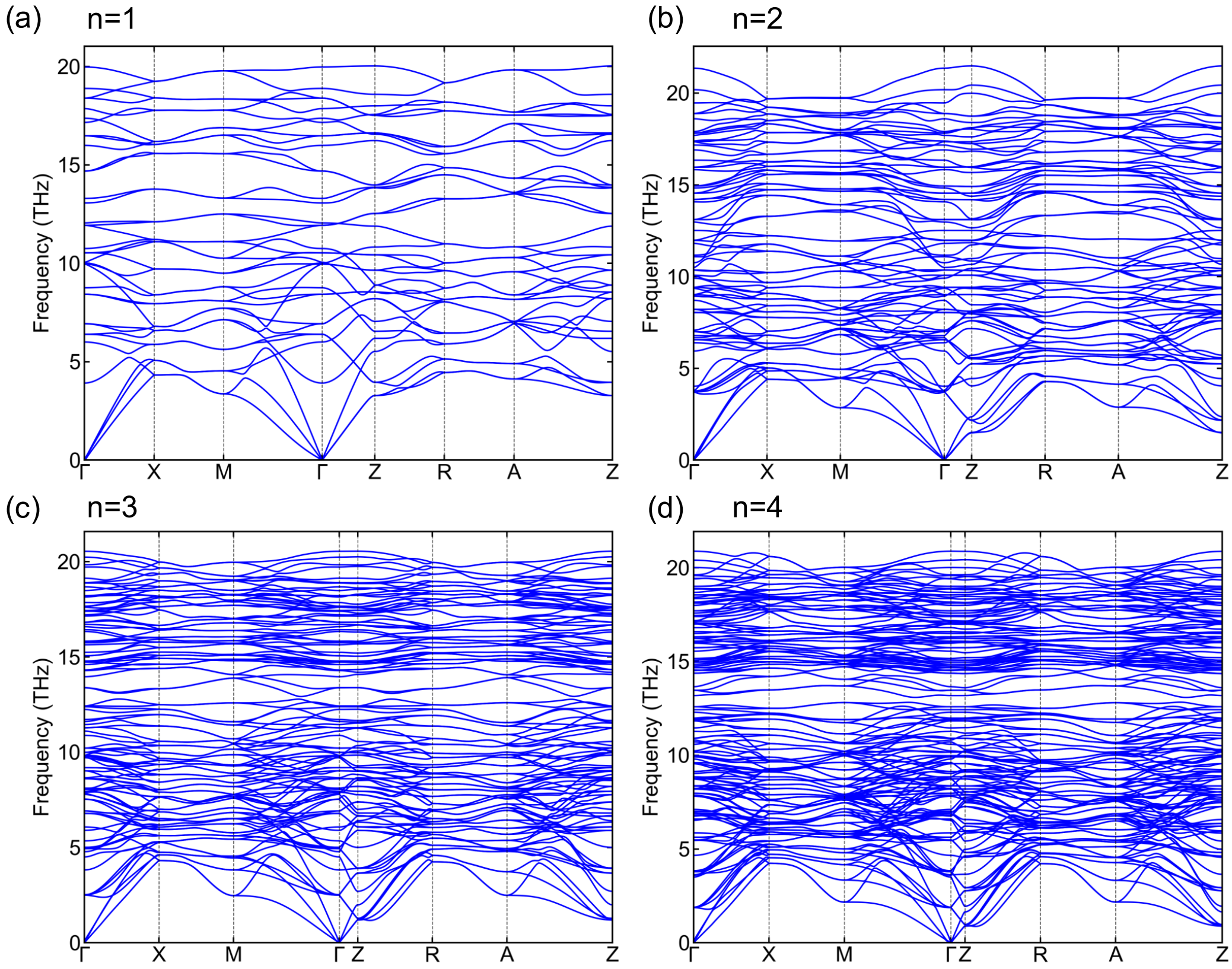

Fig. S5. Phonon dispersion spectra of $(RuO_2)_n/(MnO_2)_n$ superlattices. Calculated phonon frequencies are shown for superlattice periods of $n$=1 (a), $n$=2 (b), $n$=3 (c), and $n$=4 (d) along the high-symmetry paths of the Brillouin zone.

## Note 5. Generality of the superlattice engineering strategy

Our design strategy can be in principle generalized to a broader class of $RuO_2/MO_2$ systems to search altermagnetic candidates following specific selection criteria for the M element: (i) For structural compatibility, M should form a rutile-type oxide with a small lattice mismatch to ensure dynamical stability. (ii) For the magnetic template, M must host a stable and robust antiferromagnetic order to provide the necessary interfacial exchange field. While transition metal oxides like $IrO_2$ and $OsO_2$ satisfy the structural compatibility, they are non-magnetic and cannot stabilize the altermagnetic state of Ru. Conversely, ferromagnetic oxides such as $CrO_2$ would introduce a macroscopic net magnetization. Therefore, among rutile oxides, $MnO_2$ stands out as an ideal and unique candidate that perfectly satisfies all structural, symmetry, and magnetic coupling requirements.

To explicitly demonstrate this strict requirement and the insufficiency of structural compatibility alone, we present additional calculations on two comparative superlattices, $RuO_2/IrO_2$ and $RuO_2/OsO_2$, below. As shown in Figure S6(a), the magnetic moments of Ru in both the $RuO_2/IrO_2$ and $RuO_2/OsO_2$ superlattices behave similarly to those in bulk $RuO_2$: they remain negligible at low $U$-$J$ on Ru sites and only emerge when $U$-$J$ exceeds a critical threshold. Furthermore, the band structures of the $RuO_2/IrO_2$ and $RuO_2/OsO_2$ superlattices at $U_{\mathrm{eff}}^{\mathrm{Ru}}$= 0 eV [Fig. S6(b) and (c)] show complete spin degeneracy, confirming their non-magnetic nature. These results unequivocally underscore the scientific significance of our superlattice engineering strategy: it is the robust intrinsic antiferromagnetic order of $MnO_2$ that anchors the Ru moments through interfacial exchange coupling (magnetic proximity effect), effectively bypassing the magnetic instability inherent in bulk $RuO_2$.

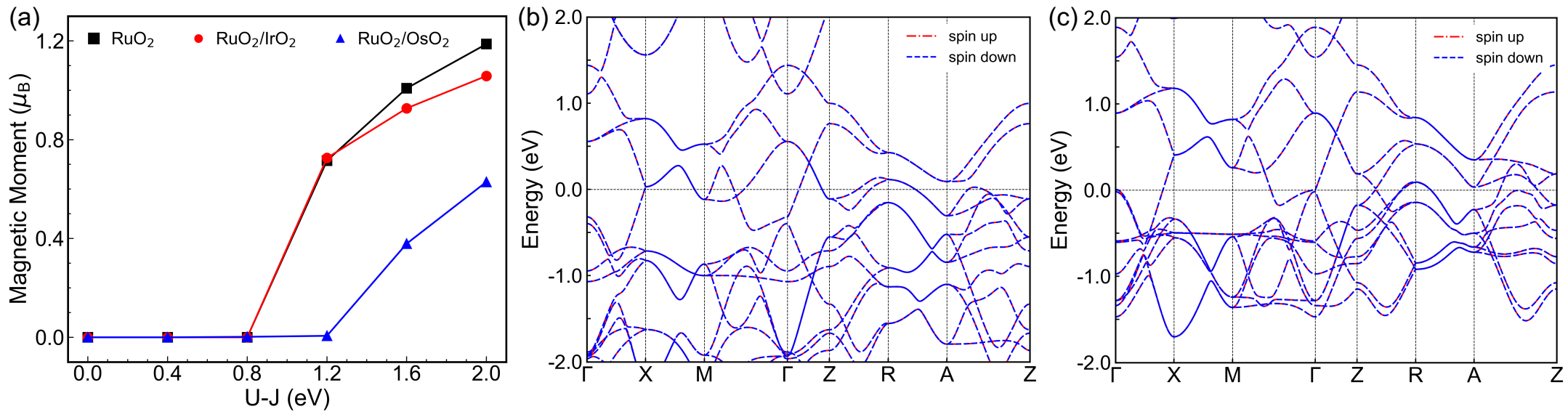

Fig. S6. (a) Calculated local magnetic moments on Ru sites as a function of the effective Hubbard interaction $U$-$J$ on Ru sites for bulk $RuO_2$ (black squares), the $RuO_2/IrO_2$ superlattice (red circles), and the $RuO_2/OsO_2$ superlattice (blue triangles). Electronic band structures of (b) the $RuO_2/IrO_2$ superlattice and (c) $RuO_2/OsO_2$ superlattice at $U_{\mathrm{eff}}^{\mathrm{Ru}}$= 0 eV, calculated without spin-orbit coupling.

## Note 6. Properties of pristine rutile $MnO_2$

Within our density functional theory framework, the collinear antiferromagnetic state is found to be the energetic ground state. At a fixed $U_{\text{eff}}^{\text{Mn}} = 3$ eV, the antiferromagnetic configuration is approximately 36 meV per Mn atom lower in energy than the ferromagnetic configuration, indicating a robust magnetic order that provides a reliable template for the superlattice.

The calculated band structure of pristine rutile $MnO_2$ is shown in Fig. S7(a). The system exhibits a semiconducting gap and hosts characteristic momentum-dependent spin splitting dictated by the rutile symmetry. In the $RuO_2/MnO_2$ superlattice, the primary contribution of the $MnO_2$ layer is to act as a magnetic anchor. By providing a stable interfacial exchange field, $MnO_2$ effectively anchors and locks the magnetic moments of the Ru atoms into a stable altermagnetic order, as shown in Table S1 in Note 3.

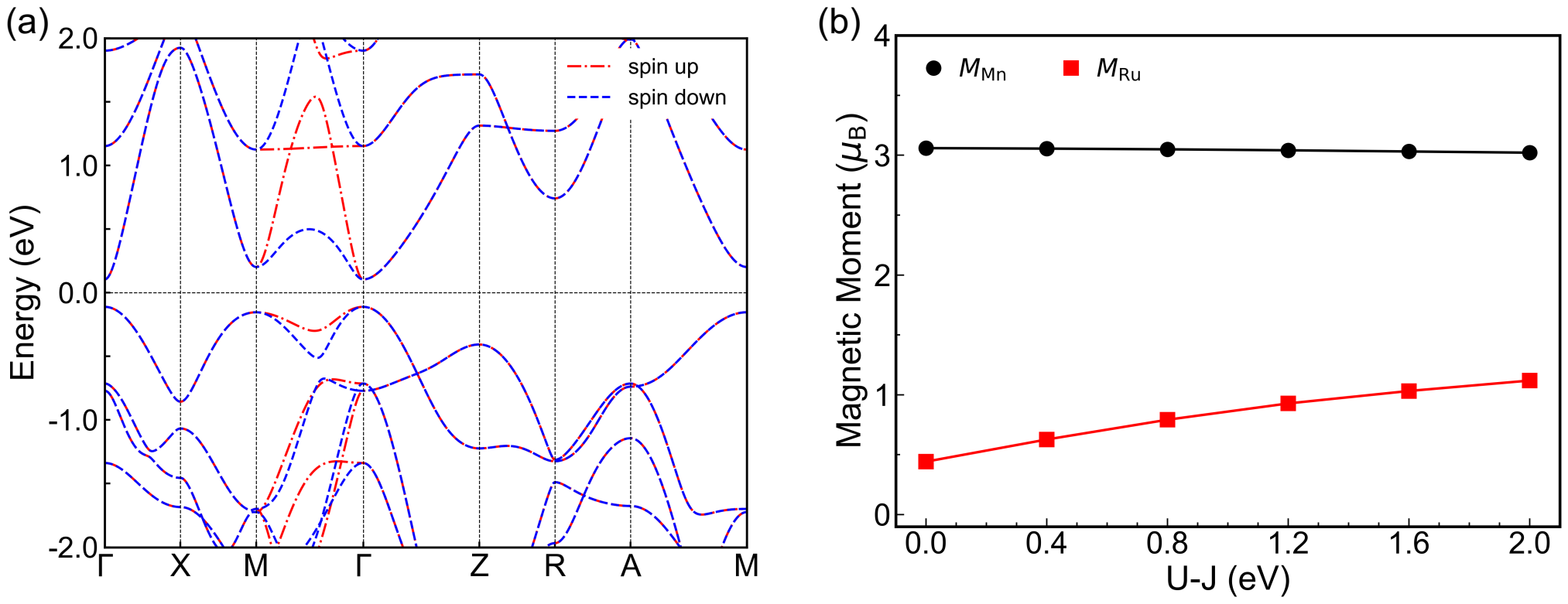

Fig. S7. (a) Electronic band structure of pristine rutile $MnO_2$ calculated without spin-orbit coupling ($U_{\text{eff}}^{\text{Mn}} = 3$eV). The system is a semiconductor with characteristic momentum-dependent spin splitting. (b) Evolution of the calculated local magnetic moments on Mn (black symbols) and Ru (red symbols) sites in the $RuO_2/MnO_2$ superlattice as a function of the effective Hubbard interaction $U_{\text{eff}}^{\text{Ru}} = U - J$.

The robustness of this magnetic template is demonstrated in Fig. S7(b), which plots the evolution of the local magnetic moments on Mn and Ru sites as a function of the Hubbard $U_{\text{eff}}^{\text{Ru}}$ parameter. While the Ru moments are sensitive to the correlation strength, the Mn moments remain stable at approximately $3.0\mu_B$ across the entire range of $U_{\text{eff}}^{\text{Ru}}$ values. This stability confirms that the $MnO_2$ layer serves as the physical origin for inducing and maintaining the $d$-wave altermagnetism in the superlattice system, independent of the variations in the Ru electronic environment.

## Note 7. Evolution of Ru 4*d* spin splitting with correlation strength

To visualize the evolution of the spin splitting with the effective Hubbard parameter, we calculated the spin-resolved Fermi-surface contours of the superlattice in the $k_z = \pi/c$ plane [Fig. S8(a)-(c)]. The results for three representative $U_{\mathrm{eff}}^{\mathrm{Ru}}$ values (0, 0.5, and 1.0 eV) all exhibit the characteristic momentum-dependent splitting of $d$-wave altermagnetism, with a progressive separation of the spin-resolved contours as $U_{\mathrm{eff}}^{\mathrm{Ru}}$ increases. In the absence of spin splitting, the spin-up (red) and spin-down (blue) contours would coincide. While spin splitting is already present at $U_{\mathrm{eff}}^{\mathrm{Ru}}$ =0 eV, substantial overlap remains between the contours in the vicinity of the Z point. As $U_{\mathrm{eff}}^{\mathrm{Ru}}$ increases to 0.5 eV, this overlapping region visibly shrinks, and at $U_{\mathrm{eff}}^{\mathrm{Ru}}$ = 1.0 eV, the red and blue contours become fully separated. This progressive reduction of the contour overlap indicates that increasing the Ru-site correlation enhances the momentum-dependent spin splitting.

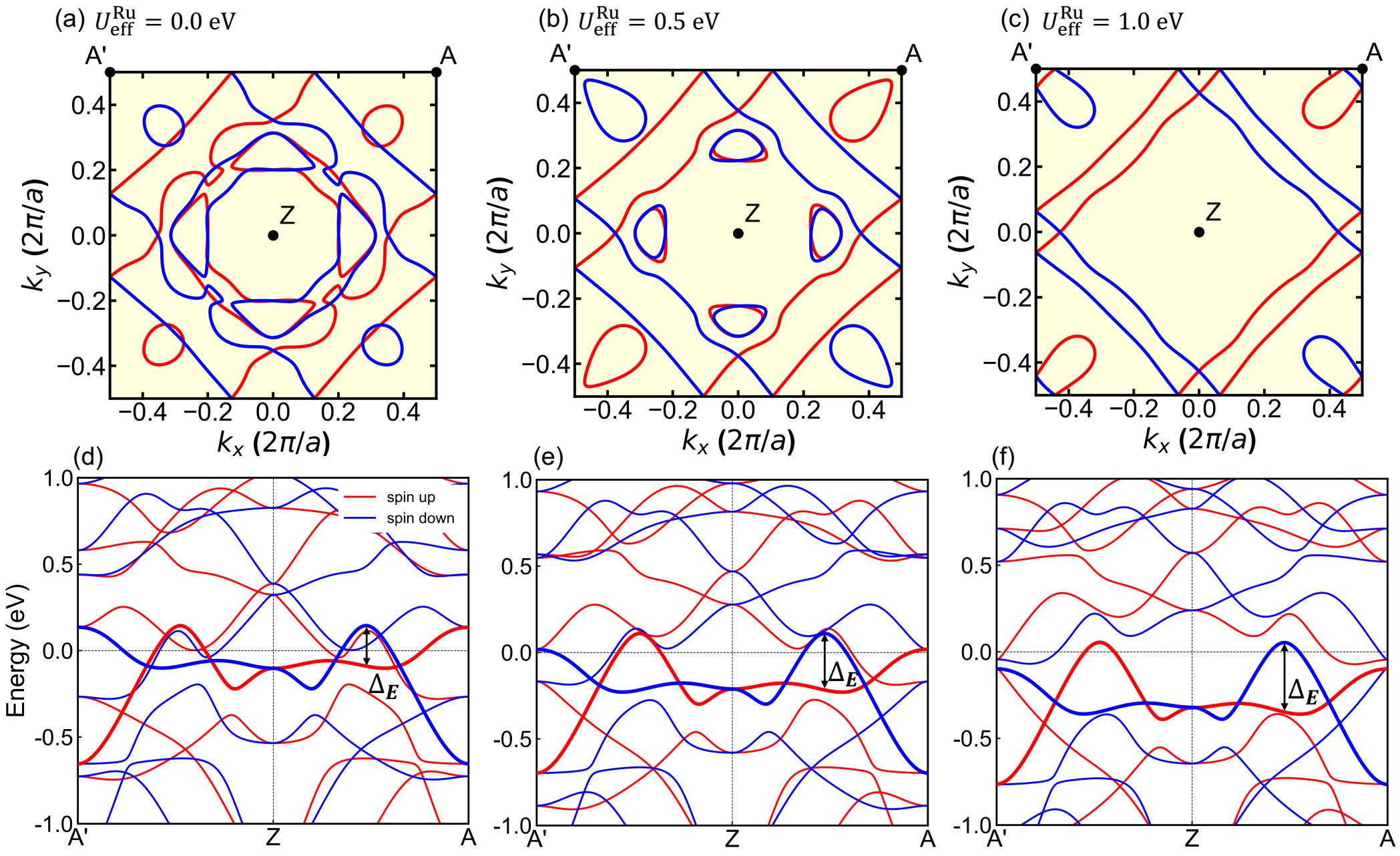

Fig. S8. (a)-(c) Spin-resolved Fermi-surface contours of the RuO2/MnO2 superlattice in the $k_z = \pi/c$ plane, calculated with $U_{\mathrm{eff}}^{\mathrm{Ru}} = 0$ eV, 0.5 eV, and 1.0 eV, respectively. (d)-(f) Corresponding spin-resolved band structures along the high-symmetry Z-A path. A representative pair of spin-resolved bands is highlighted by thicker lines and the vertical arrows indicate their energy separation $\Delta_E$. All calculations were performed without spin-orbit coupling to isolate the exchange-driven spin splitting. The red and blue solid lines represent the spin-up and spin-down states, respectively.

To provide a more direct and quantitative visualization of this energy separation, we also present the corresponding spin-resolved band structures along the high-symmetry Z-A path [Fig. S8(d)-(f)]. Here, we focus on a representative pair of spin-resolved bands (highlighted by thicker lines) that are degenerate at the high-symmetry Z point, where the effective altermagnetic exchange field vanishes by symmetry. Following these bands away from Z, we extract their energy separation $\Delta_{\boldsymbol{E}}$. As $U_{\mathrm{eff}}^{\mathrm{Ru}}$ increases from 0 to 0.5 and 1.0 eV, the splitting increases monotonically, with $\Delta_{\boldsymbol{E}}$=0.23, 0.33, and 0.40 eV, respectively. These quantitative results demonstrate that increasing the Ru-site correlation progressively enhances the momentum-dependent spin splitting.

**Note 8. Symmetry elements and allowed spin conductivity responses**

To systematically summarize the symmetry constraints on the spin transport properties, we present the complete symmetry compatible spin conductivity components in Table S2. Checkmark and cross symbols are used to denote whether a specific tensor component is allowed or forbidden by a given symmetry operation. Furthermore, we have introduced a color-shading scheme to highlight the symmetry operations that remain preserved under the three specific crystallographic configurations discussed in the main text.

Table S2. Symmetry constraints on the spin conductivity tensor components in the superlattice. The rows list the relevant tensor components grouped by their indices and their time-reversal (T) parity. The columns represent the non-trivial magnetic symmetry operations. Checkmarks (✓) and crosses (×) indicate whether a specific component is allowed or strictly forbidden by a given symmetry operation, respectively. The background colors highlight the symmetry operations preserved in Configuration (i) (yellow), Configuration (ii) (blue), and Configuration (iii) (green). The two columns on the far right denote additional effective symmetry operations that do not belong to the parent magnetic space group but specifically emerge in Configuration (iii).

| | $\{C_{2x}\|\tau\}$ | $\{C_{2y}\|\tau\}$ | $C_{2z}$ | $TS_{4z}^{-}$ | $TS_{4z}^{+}$ | $T\{M_{x=y}\|\tau\}$ | $T\{M_{x=-y}\|\tau\}$ | $T\{M_x\|\tau\}$ | $T\{M_y\|\tau\}$ |
|---|---|---|---|---|---|---|---|---|---|
| $\sigma_{xy}^{\mathrm{odd},z}$ | ✓ | ✓ | ✓ | ✓ | ✓ | ✓ | ✓ | × | × |
| $\sigma_{xy}^{\mathrm{even},z}$ | ✓ | ✓ | ✓ | ✓ | ✓ | ✓ | ✓ | ✓ | ✓ |
| $\sigma_{xz}^{\mathrm{odd},y}$ | ✓ | ✓ | ✓ | ✓ | ✓ | ✓ | ✓ | × | × |
| $\sigma_{xz}^{\mathrm{even},y}$ | ✓ | ✓ | ✓ | ✓ | ✓ | ✓ | ✓ | ✓ | ✓ |
| $\sigma_{yz}^{\mathrm{odd},x}$ | ✓ | ✓ | ✓ | ✓ | ✓ | ✓ | ✓ | × | × |
| $\sigma_{yz}^{\mathrm{even},x}$ | ✓ | ✓ | ✓ | ✓ | ✓ | ✓ | ✓ | ✓ | ✓ |
| $\sigma_{xy}^{\mathrm{odd},x}$ | × | ✓ | × | ✓ | ✓ | ✓ | ✓ | ✓ | × |
| $\sigma_{xy}^{\mathrm{even},x}$ | × | ✓ | × | ✓ | ✓ | ✓ | ✓ | × | ✓ |
| $\sigma_{xz}^{\mathrm{odd},x}$ | × | × | ✓ | ✓ | ✓ | ✓ | ✓ | ✓ | ✓ |
| $\sigma_{xz}^{\mathrm{even},x}$ | × | × | ✓ | ✓ | ✓ | ✓ | ✓ | × | × |
| $\sigma_{yx}^{\mathrm{odd},y}$ | ✓ | × | × | ✓ | ✓ | ✓ | ✓ | × | ✓ |
| $\sigma_{yx}^{\mathrm{even},y}$ | ✓ | × | × | ✓ | ✓ | ✓ | ✓ | ✓ | × |
| $\sigma_{yz}^{\mathrm{odd},y}$ | × | × | ✓ | ✓ | ✓ | ✓ | ✓ | ✓ | ✓ |
| $\sigma_{yz}^{\mathrm{even},y}$ | × | × | ✓ | ✓ | ✓ | ✓ | ✓ | × | × |
| $\sigma_{zx}^{\mathrm{odd},z}$ | ✓ | × | × | ✓ | ✓ | ✓ | ✓ | × | ✓ |
| $\sigma_{zx}^{\mathrm{even},z}$ | ✓ | × | × | ✓ | ✓ | ✓ | ✓ | ✓ | × |
| $\sigma_{zy}^{\mathrm{odd},z}$ | × | ✓ | × | ✓ | ✓ | ✓ | ✓ | ✓ | × |
| $\sigma_{zy}^{\mathrm{even},z}$ | × | ✓ | × | ✓ | ✓ | ✓ | ✓ | × | ✓ |
| $\sigma_{xx}^{\mathrm{odd},y}$ | × | ✓ | × | ✓ | ✓ | ✓ | ✓ | ✓ | × |
| $\sigma_{xx}^{\mathrm{even},y}$ | × | ✓ | × | ✓ | ✓ | ✓ | ✓ | × | ✓ |
| $\sigma_{xx}^{\mathrm{odd},z}$ | × | × | ✓ | ✓ | ✓ | ✓ | ✓ | ✓ | ✓ |
| $\sigma_{xx}^{\mathrm{even},z}$ | × | × | ✓ | ✓ | ✓ | ✓ | ✓ | × | × |
| $\sigma_{yy}^{\mathrm{odd},x}$ | ✓ | × | × | ✓ | ✓ | ✓ | ✓ | × | ✓ |
| $\sigma_{yy}^{\mathrm{even},x}$ | ✓ | × | × | ✓ | ✓ | ✓ | ✓ | ✓ | × |
| $\sigma_{yy}^{\mathrm{odd},z}$ | × | × | ✓ | ✓ | ✓ | ✓ | ✓ | ✓ | ✓ |
| $\sigma_{yy}^{\mathrm{even},z}$ | × | × | ✓ | ✓ | ✓ | ✓ | ✓ | × | × |
| $\sigma_{zz}^{\mathrm{odd},x}$ | ✓ | × | × | ✓ | ✓ | ✓ | ✓ | × | ✓ |
| $\sigma_{zz}^{\mathrm{even},x}$ | ✓ | × | × | ✓ | ✓ | ✓ | ✓ | ✓ | × |
| $\sigma_{zz}^{\mathrm{odd},y}$ | × | ✓ | × | ✓ | ✓ | ✓ | ✓ | ✓ | × |
| $\sigma_{zz}^{\mathrm{even},y}$ | × | ✓ | × | ✓ | ✓ | ✓ | ✓ | × | ✓ |

## Note 9. Isolation of altermagnetic signals from parasitic ferromagnetic artifacts

The symmetry-based transport criterion is capable of isolating any artifacts due to uncompensated or canted magnetic moments at the interfaces. We classify the components of these magnetic moments into the parallel and perpendicular ones compared with the Néel order in the superlattice. The perpendicular moments would not affect any results in our designed TMR measurement scheme: because we need to detect the spin polarization collinear with the Néel order, the magnetization in the ferromagnetic layer is supposed to align with the Néel-order direction. Thus, the measured TMR would not be affected by any spin currents generated by the uncompensated or canted moments perpendicular to the Néel order.

The parallel component collinear with the Néel order (assuming along $z$) leads to the symmetric spin conductivity with current along $[110]$ ($x$) and $[\bar{1}10]$ ($y$) directions, giving $\sigma_{xx}^{z} = \sigma_{yy}^{z}$, while the intrinsic $d$-wave altermagnetic state (governed by the $TS_{4z}$ symmetry operation) strictly dictates an antisymmetric relation: $\sigma_{xx}^{z,AM} = -\sigma_{yy}^{z,AM}$ [as explicitly demonstrated by our calculations in Fig. 3(d)]. Therefore, one simply needs to measure the longitudinal spin-polarized current along the $[110]$ and $[\bar{1}10]$ directions. By extracting the antisymmetric difference $\frac{1}{2}(\sigma_{xx}^{z} - \sigma_{yy}^{z})$, any $z$-polarized ferromagnetic artifacts are mathematically canceled out, unambiguously isolating the pure altermagnetic contribution.

In summary, the proposed superlattice platform, combined with our symmetry-based transport criterion, inherently possesses a highly robust, multi-tiered mechanism to completely isolate the intrinsic altermagnetic signal from any such parasitic artifacts. Our design provides a definitive, airtight experimental roadmap to decouple intrinsic altermagnetism from parasitic signals.

**Note 10. Influence of local defects on transport signatures**

In realistic oxide superlattices, imperfections such as interfacial cation intermixing and point defects may occur and could influence the transport properties. To quantitatively evaluate the robustness of our proposed transport criterion against such imperfections, we performed additional first-principles calculations incorporating representative local defect structures.

For the computational model, we constructed an in-plane 2×2 supercell of the $n$=1 superlattice. This choice provides a stringent test of disorder robustness because a single defect in this minimal structure corresponds to a relatively high effective defect concentration. As illustrated in Fig. S9(a), we considered four representative local defect structures (labeled D1–D4), including two inequivalent neighboring Ru–Mn antisite defects (D1 and D2) to model interfacial cation intermixing and two inequivalent oxygen vacancies (D3 and D4). We then calculated the key T-odd spin conductivity components $\sigma_{xz}^{x}$ and $\sigma_{xx}^{z}$ in configuration (iii), which are the central transport quantities used in our symmetry-based criterion for identifying the altermagnetic response.

As shown in Fig. S9(b), these transport quantities retain a sizable magnitude and consistent sign across the considered defect structures, with the variations among different defects reflected by the error bars. These results indicate that the proposed symmetry-based transport criterion remains robust against the representative local perturbations examined here and does not require an ideal defect-free structure.

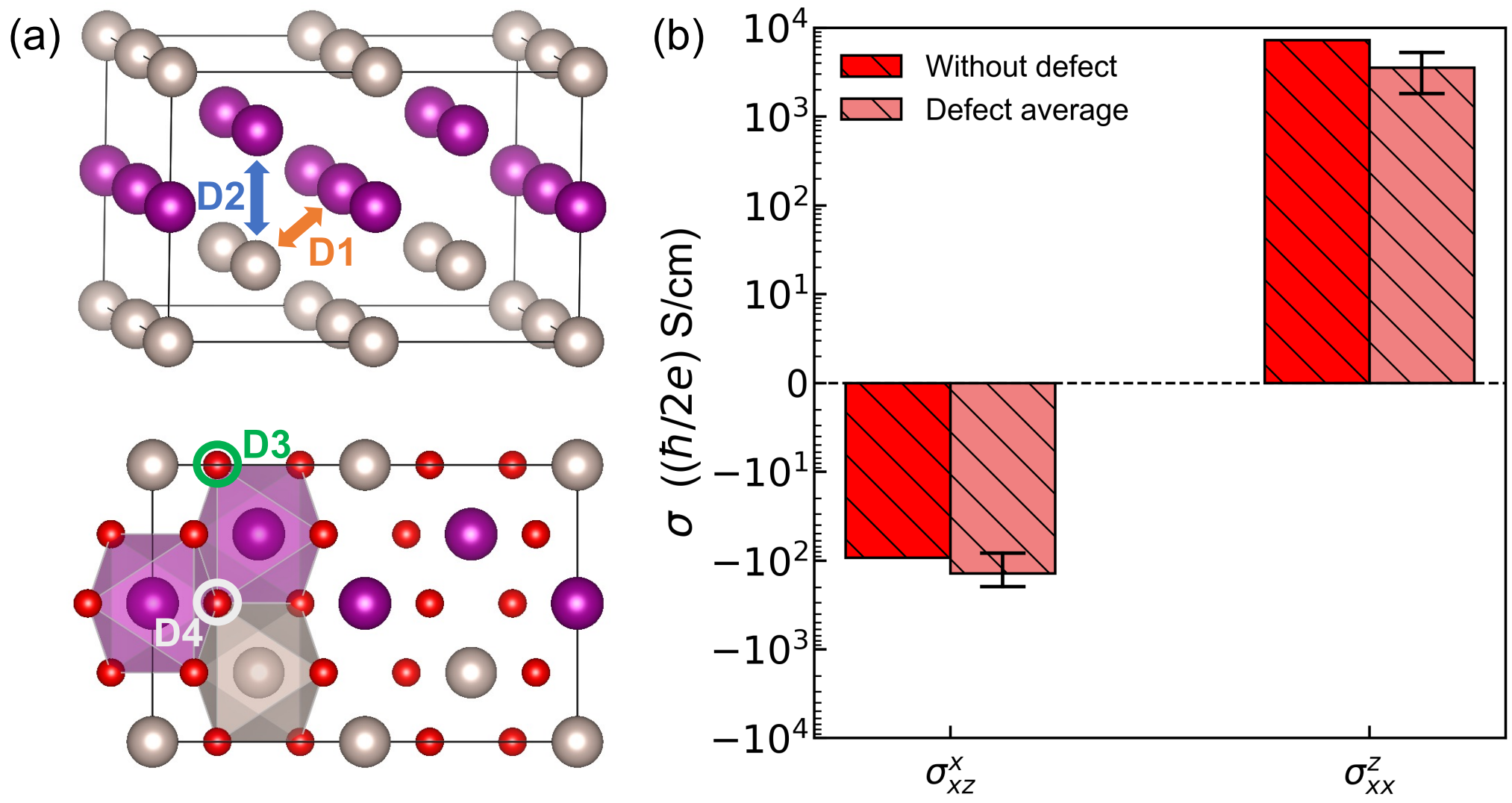

Fig. S9. (a) Schematic illustrations of the four representative local defect structures (D1–D4) in an in-plane 2 × 2 supercell of the $n$=1 superlattice. Gray, purple, and small red spheres represent Ru, Mn, and O atoms, respectively. The upper panel omits O atoms for clarity. The orange (D1) and blue (D2) double-headed arrows indicate two inequivalent neighboring Ru–Mn antisite defects. In the lower panel, the green (D3) and white (D4) circles indicate the sites of two inequivalent oxygen vacancies. (b) Calculated spin conductivity tensor components $\sigma^x_{xz}$ and $\sigma^z_{xx}$ in configuration (iii) for the pristine $n$=1 superlattice and the average over four defect structures. Error bars indicate the variation among the four defect structures.

## References

[1] G. Kresse and J. Furthmüller, Efficient iterative schemes for *ab initio* total-energy calculations using a plane-wave basis set, Phys. Rev. B **54**, 11169 (1996).

[2] G. Kresse and D. Joubert, From ultrasoft pseudopotentials to the projector augmented-wave method, Phys. Rev. B **59**, 1758 (1999).

[3] P. E. Blöchl, Projector augmented-wave method, Phys. Rev. B **50**, 17953 (1994).

[4] J. P. Perdew, K. Burke, and M. Ernzerhof, Generalized gradient approximation made simple, Phys. Rev. Lett. **77**, 3865 (1996).

[5] S. L. Dudarev, G. A. Botton, S. Y. Savrasov, C. J. Humphreys, and A. P. Sutton, Electron-energy-loss spectra and the structural stability of nickel oxide: An LSDA+U study, Phys. Rev. B **57**, 1505 (1998).

[6] Y. Noda, K. Ohno, and S. Nakamura, Momentum-dependent band spin splitting in semiconducting $MnO_2$: a density functional calculation, Phys. Chem. Chem. Phys. **18**, 13294 (2016).

[7] M. Chepkoech, D. P. Joubert, and G. O. Amolo, First principles calculations of the thermoelectric properties of $\alpha$-$MnO_2$ and $\beta$-$MnO_2$, Eur. Phys. J. B **91**, 301 (2018).

[8] A. Gauswami and P. K. Jha, Strain-controlled chiral magnons and spin-dependent Seebeck effect in tetragonal $\beta$-$MnO_2$ altermagnet: a DFT study, J. Phys.: Condens. Matter **38**, 025802 (2026).

[9] Z. Qian, Y. Yang, S. Liu, and C. Wu, Fragile unconventional magnetism in $RuO_2$ by proximity to Landau-Pomeranchuk instability, Phys. Rev. B **111**, 174425 (2025).

[10] A. A. Mostofi, J. R. Yates, G. Pizzi, Y.-S. Lee, I. Souza, D. Vanderbilt, and N. Marzari, An updated version of wannier90: A tool for obtaining maximally-localised Wannier functions, Comput. Phys. Commun. **185**, 2309 (2014).